%% file: main.tex
\documentclass[ALICE,manyauthors]{cernphprep}
\usepackage[comma,square,numbers,sort&compress]{natbib}
\usepackage{hyperref}
\usepackage{lineno}
\usepackage{xspace}
\usepackage{xcolor}
\begin{document}
\input{commands.tex}
\newcommand{\jpsi} {\ensuremath{{\mathrm J}/\psi}\xspace}
\newcommand{\psip} {\ensuremath{\psi'}\xspace}
\newcommand{\AD}   {\rm{AD}\xspace}
\newcommand{\ADA}  {\rm{ADA}\xspace}
\newcommand{\ADC}  {\rm{ADC}\xspace}
\newcommand{\fI}   {f_{\rm I}}
\newcommand{\fd}   {f_{\rm D}}
\newcommand\todo[1]{\textcolor{red}{#1}}
\newcommand{\starlight}{STARlight\xspace}

\begin{titlepage}
\PHyear{2019}       
\PHnumber{069}      
\PHdate{08 April}  

\title{Coherent \jpsi photoproduction at forward rapidity in ultra-peripheral Pb--Pb collisions at \fivenn}
\ShortTitle{Coherent \jpsi photoproduction at forward rapidity in Pb-Pb UPC}   

\Collaboration{ALICE Collaboration\thanks{See Appendix~\ref{app:collab} for the list of collaboration members}}
\ShortAuthor{ALICE Collaboration} 

\begin{abstract}

The ALICE collaboration performed the first rapidity-differential measurement of coherent \jpsi photoproduction in ultra-peripheral \PbPb collisions  at a center-of-mass energy \fivenn.
The \jpsi is detected via its dimuon decay in the forward rapidity region ($-4.0 < y < -2.5$) for events where the hadronic activity is required to be minimal. The analysis is based on an event sample corresponding to an integrated luminosity of about 750~$\mu$b$^{-1}$. The cross section for coherent \jpsi production is presented in six rapidity bins. The results are compared with theoretical models for coherent \jpsi photoproduction. These comparisons indicate that gluon shadowing effects play a role in the photoproduction process. The ratio of \psip to \jpsi coherent photoproduction cross sections was measured and found to be consistent with that measured for photoproduction off protons.
\end{abstract}
\end{titlepage}

\setcounter{page}{2} 


\section{Introduction} 

Ultra-peripheral collisions (UPC) between two Pb nuclei, in which the impact parameter is larger than the sum of their radii, provide a useful way to study photonuclear reactions~\cite{Krauss:1997vr,Bertulani:2005ru,Baltz:2007kq,Contreras:2015dqa}. Photoproduction of vector mesons in these collisions has an easily identifiable experimental signature: the decay products of the vector meson, in the case of this analysis a $\mu^+\mu^-$ pair, are the only signals in an otherwise empty detector. 
This process is akin to exclusive vector meson production in electron--proton collisions, already studied extensively at HERA~\cite{Newman:2013ada}. The exchange photon, which carries a momentum transfer squared $Q^{2}$, is typified  by very small values of $Q^{2}$, and may be described as quasi-real. The intensity of the photon flux scales as the square of nuclear charge resulting in large cross sections for the photoproduction of vector mesons in \PbPb collisions at the CERN Large Hadron Collider (LHC), where the measurement presented in this Letter was performed.

Photoproduction of vector mesons on nuclei can be either coherent, where the photon couples coherently to the nucleus as a whole, or incoherent, where the photon couples to a single nucleon~\cite{Bertulani:2005ru}. Coherent production is characterized by low vector meson transverse momentum ($\langle \pt \rangle \simeq 60$ \MeVc) and by the target nucleus not breaking up. Incoherent production, corresponding to quasi-elastic scattering off a single nucleon, is characterized by a somewhat higher average transverse momentum ($\langle \pt \rangle \simeq 500$ \MeVc). The target nucleus normally breaks up in the incoherent production, but, except for single nucleons or nuclear fragments in the very forward region, no other particles are produced. The incoherent production can be accompanied by the excitation and dissociation of the target nucleon resulting in even higher transverse momenta of the produced vector mesons, extending well above 1 \GeVc~\cite{Guzey:2018tlk}.

Coherent photoproduction of the $\jpsi$ meson, a charm-anticharm bound state, is of particular interest since, for a leading order QCD calculation~\cite{Ryskin:1992ui}, its cross section is expected to scale as the square of the gluon parton density function (PDF) in the target hadron. The mass of the charm quark provides an energy scale large enough to allow for perturbative QCD calculations. For this process, a variable corresponding to Bjorken-$x$ can be defined using the mass of the vector meson ($m_{\jpsi}$) and its rapidity ($y$) as $x=(m_{\jpsi}/\sqrt{s_{\rm NN}})\exp(\pm y)$. Though next-to-leading order effects and scale uncertainties complicate extraction of gluon PDFs from $\jpsi$ photoproduction data~\cite{Jones:2015nna}, the related uncertainties are expected to largely cancel in the ratio of coherent photoproduction cross sections off nuclei and off protons~\cite{Guzey:2013qza}. Thus, coherent $\jpsi$ photoproduction off lead nuclei ($\gamma + {\rm Pb} \to \jpsi  + {\rm Pb}$) provides a powerful tool to study poorly known gluon shadowing effects at low Bjorken-$x$ values ranging from $x \sim 10^{-5}$ to $x \sim 10^{-2}$ at LHC energies~\cite{Guzey:2013xba,Contreras:2016pkc}.

The ALICE collaboration has pioneered the study of charmonium photoproduction in ultra-peripheral \PbPb collisions at the LHC at a center-of-mass energy per nucleon pair \twosevensixnn~\cite{Abelev:2012ba,Abbas:2013oua,Adam:2015sia}. Coherent \jpsi\ photoproduction was studied both at forward rapidity ($-3.6 < y < -2.6$) with the ALICE muon spectrometer and at mid-rapidity ($|y| < 0.9$) with the central barrel.
The CMS collaboration studied coherent \jpsi photoproduction accompanied by neutron emission in the semi-forward rapidity range $1.8 < |y| < 2.3$~\cite{Khachatryan:2016qhq}. The ALICE and CMS results on \jpsi\ photoproduction were compared with predictions from models available at that time, and suggested that moderate shadowing in the nucleus was necessary to describe the measurements. In particular, the nuclear gluon shadowing factor
$R_{g}$, i.e. the ratio of the nuclear gluon density distribution to the proton gluon distribution, was extracted from the ALICE measurements~\cite{Guzey:2013xba}, and found to be, at the scale of the charm quark mass, $R_g(x\sim 10^{-3}) = 0.61^{+0.05}_{-0.04}$ and $R_g(x\sim 10^{-2}) = 0.74^{+0.11}_{-0.12}$.
The ALICE collaboration also measured the coherent cross section for \psip photoproduction at mid-rapidity, and the results supported, within the experimental uncertainties, the moderate-shadowing scenario~\cite{Adam:2015sia}.

In this Letter, we present the first measurement of the coherent \jpsi photoproduction in ultra-peripheral \PbPb collisions  at a center-of-mass energy per nucleon pair \fivenn. The measurement was performed with the ALICE muon spectrometer covering the rapidity range $-4.0 < y <-2.5$. The results presented here are based on data taken in 2015 and in 2018, during  Run~2 of the LHC. The recorded data sample is some 200 times larger than the data used in the \twosevensixnn \PbPb analysis~\cite{Abelev:2012ba}. The new result is based on the absolute luminosity normalization in contrast to previous measurement based on the normalization relative to the continuum $\gamma \gamma \to \mu^+ \mu^-$ cross section predicted by \starlight~\cite{Klein:1999qj}. These two improvements imply a considerable reduction in the statistical and systematic uncertainties and the possibility to study the rapidity dependence in the forward region.

\section{Detector description}

The ALICE detector and its performance are described in~\cite{Aamodt:2008zz,Abelev:2014ffa}. Muons from \jpsi decays are measured in the single-arm muon spectrometer, while other activity is vetoed using the Silicon Pixel Detector (\SPD),  the \VZERO and ALICE Diffractive (\AD) detectors. The muon spectrometer covers the pseudorapidity interval $-4.0<\eta<-2.5$. It consists of a ten interaction length absorber followed by five tracking stations, the third of which is placed inside a dipole magnet with a 3 T$\cdot$m integrated magnetic field, a 7.2 interaction length iron wall, and a trigger system located downstream of the iron wall. Each tracking station is made of two planes of cathode pad chambers, while the trigger system consists of four planes of resistive plate chambers arranged in two stations. Muon tracks are reconstructed using the tracking algorithm described in~\cite{Aamodt:2011gj}.
The central region $|\eta|<$ 1.4 is covered by the \SPD consisting of two cylindrical layers of silicon pixel sensors. The \VZERO detector is composed of the \VZEROA and \VZEROC sub-detectors, consisting of 32 cells each and covering the pseudorapidity interval $2.8< \eta <5.1$ and $-3.7<\eta <-1.7$, respectively. The newly installed AD detector is composed of the \ADC and \ADA sub-detectors located at $-19.5$ and $+16.9$ m from the interaction point covering the pseudorapidity ranges $-7.0 < \eta < -4.9$ and $4.7 <\eta < 6.3$, respectively~\cite{N.Cartiglia:2015gve}. The \VZERO and \AD detectors are scintillator tile arrays with a time resolution better than 1 ns, allowing one to distinguish between beam-beam and beam-gas interactions. 

\section{Data analysis}

The analysis presented in this publication is based on a sample of events collected during the 2015 and 2018 \PbPb data taking periods at \fivenn, characterized by similar beam conditions and interaction rates. The muon spectrometer performance was stable during the whole Run~2 thus allowing for the merging of the two data sets.
The trigger required two oppositely charged tracks in the muon spectrometer, and vetoes on \VZEROA, \ADA and \ADC beam-beam interactions. The single muon trigger threshold was set to a transverse momentum $\pt = 1$ \GeVc ~\cite{Bossu:2012jt}. The integrated luminosities of 216 $\mu$b$^{-1}$ in 2015 and 538 $\mu$b$^{-1}$ in 2018, with relative systematic uncertainty of 5\%, were estimated from the counts of a reference trigger, based on multiplicity selection in the \VZERO detector. The reference trigger cross section was derived from Glauber-model-based estimates of the inelastic \PbPb cross section~\cite{Loizides:2017ack}. 

Events with only two tracks with opposite electric charge (unlike-sign) in the muon spectrometer were selected offline. The pseudorapidity of each track was required to be within the range $-4.0<\eta<-2.5$. 
The tracks had to fulfill the requirements, described in~\cite{Abelev:2012ba}, on the radial coordinate of the track at the end of the absorber and on the extrapolation to the nominal vertex. Track segments in the tracking chambers had to be matched with corresponding segments in the trigger chambers. 

Additional offline vetoes on the V0A, ADA and ADC detector signals were applied to ensure the exclusive production of the muon pair. Exclusivity in the muon spectrometer region was assured by requiring a maximum of 2 fired cells in \VZEROC. Online and offline veto requirements may result in significant inefficiencies (denoted as veto inefficiencies) in the exclusive \jpsi cross section measurements due to additional \VZERO and \AD detector activity induced by independent hadronic or electromagnetic pile--up processes accompanying the coherent \jpsi photoproduction. The probability of hadronic pile--up did not exceed $0.2\%$, however there was a significant pile--up contribution from the electromagnetic electron pair production process $\gamma \gamma \to e^+ e^-$. The veto inefficiency induced by these pile--up effects in the \VZEROA, \VZEROC, \ADA and \ADC detectors, was estimated using the events selected with an unbiased trigger based only on the timing of bunches crossing the interaction region. The veto rejection probability, defined as the probability to detect activity in these sub-detectors, was found to scale linearly with the expected number of collisions per bunch crossing reaching 10\% in \VZEROA. The veto inefficiency correction factors were determined by weighting the corresponding veto rejection probabilities over periods with different pile--up conditions, taking the luminosity of each period as a weight. The veto inefficiency of the \VZEROA online and offline selection was found to be $p_{\rm \VZEROA} = (4.6 \pm  0.2)\%$, where the uncertainty is related to the limited statistics in the unbiased trigger sample. The veto inefficiencies in \ADA ($p_{\rm \ADA}$) and \ADC ($p_{\rm \ADC}$) were found to be about $0.2\%$, because these detectors are far away from the interaction point and are thus much less affected by soft $\mathrm{e}^+ \mathrm{e}^-$ pairs. The veto inefficiency in \VZEROC, associated with the requirement of maximum 2 fired cells, was found to be negligible. The average veto efficiency correction factor $\epsilon_{\rm veto} = 95.0\%$, and this is applied to raw \jpsi yields to account for hadronic and electromagnetic pile--up processes, was calculated as a product of individual veto inefficiencies $\epsilon_{\rm veto} = (1-p_{\rm \VZEROA})(1-p_{\rm \ADA})(1-p_{\rm \ADC})$. 

The acceptance and efficiency of \jpsi and \psip reconstruction were evaluated using a large sample of coherent and incoherent \jpsi and \psip events generated by \starlight 2.2.0~\cite{Klein:2016yzr} with decay muons tracked in a model of the apparatus implemented in GEANT 3.21~\cite{Brun:1994aa}. The model includes a realistic description of the detector performance during data taking as well as its variation with time. The acceptance and efficiency of feed-down $\psip \to \jpsi + \pi\pi$ decays were also evaluated using the \starlight generator under the assumption that feed-down $\jpsi$ mesons inherit the transverse polarization of their \psip parents, as indicated by previous measurements~\cite{Bai:1999mj}. The same samples were also used for modelling the signal shape and different background contributions.

\begin{figure}[tb]
\begin{center}
\includegraphics[width = 0.45\textwidth]{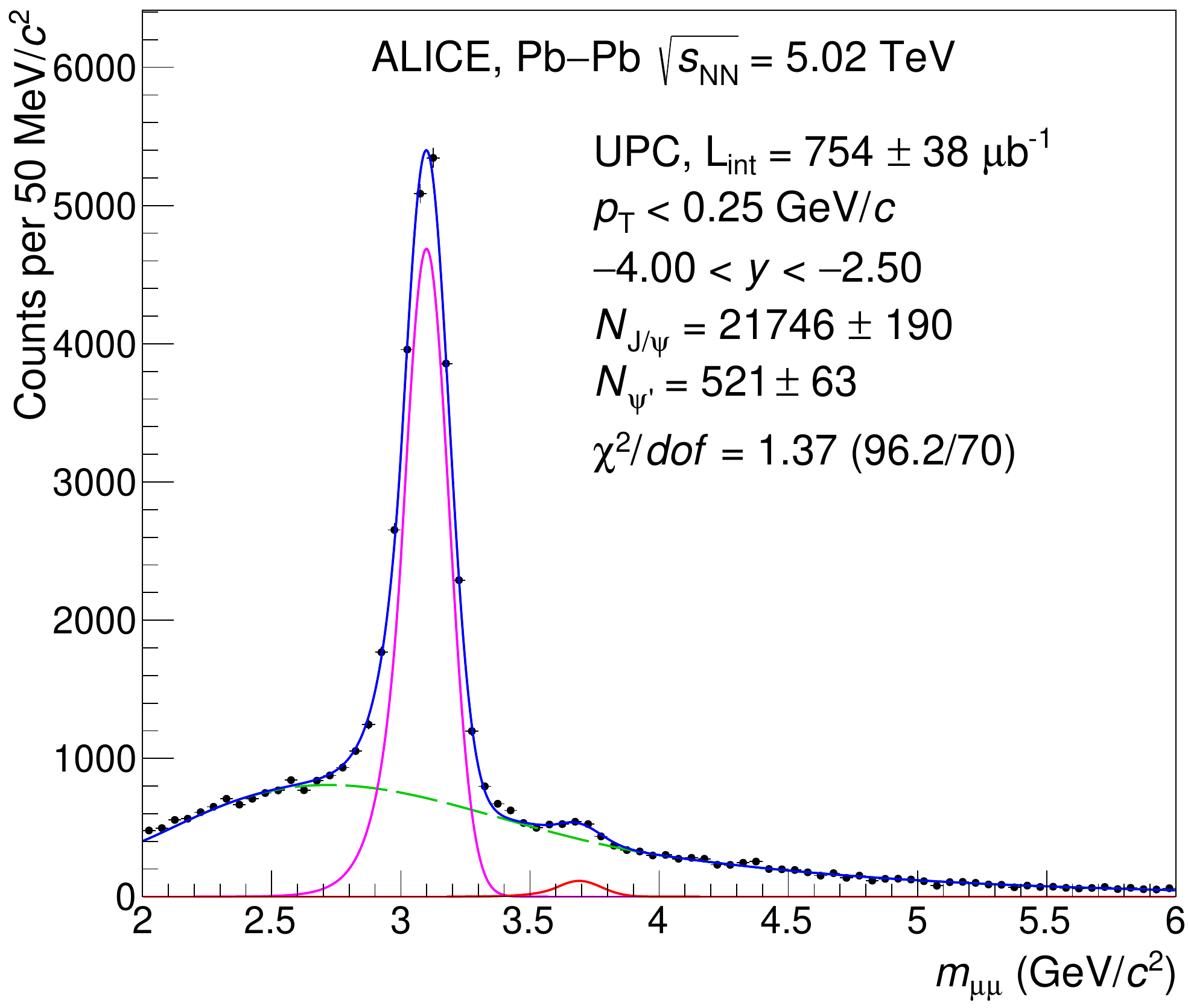}
\quad
\includegraphics[width = 0.45\textwidth]{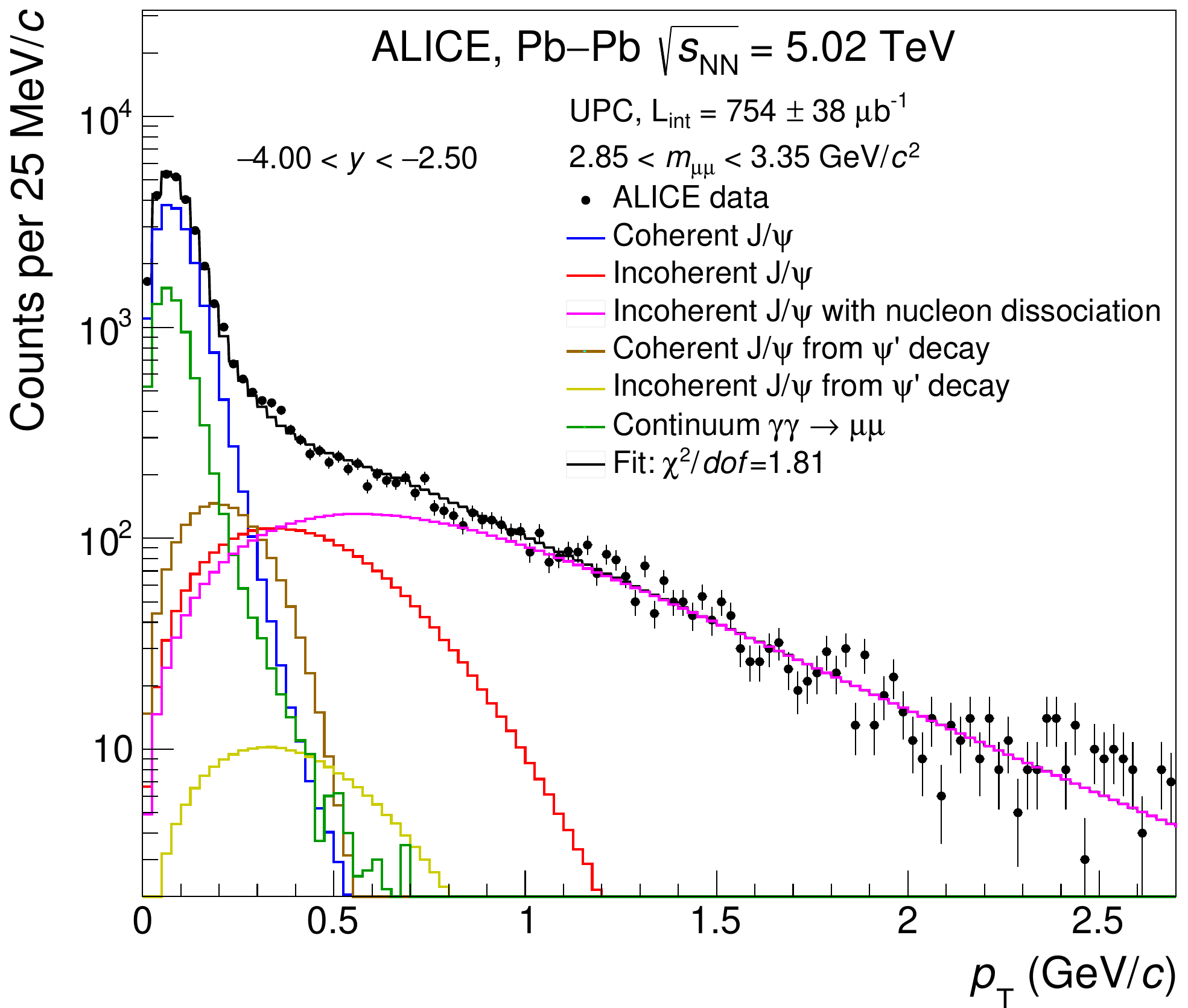}
\end{center}
\caption{
Left: (color online) Invariant mass distribution for muon pairs satisfying the event selection described in the text. The dashed green line corresponds to the background.  The solid magenta and red lines correspond to Crystal Ball functions representing \jpsi and \psip signals, respectively. The solid blue line corresponds to the sum of background and signal functions. Right: transverse momentum distribution for muon pairs in the range $2.85 < m_{\mu\mu} < 3.35$ GeV/$c^{2}$(around the \jpsi mass).}
\label{fig:m0pt0}
\end{figure}

\begin{figure}[h!]
\begin{center}
\includegraphics[width = 0.45\textwidth]{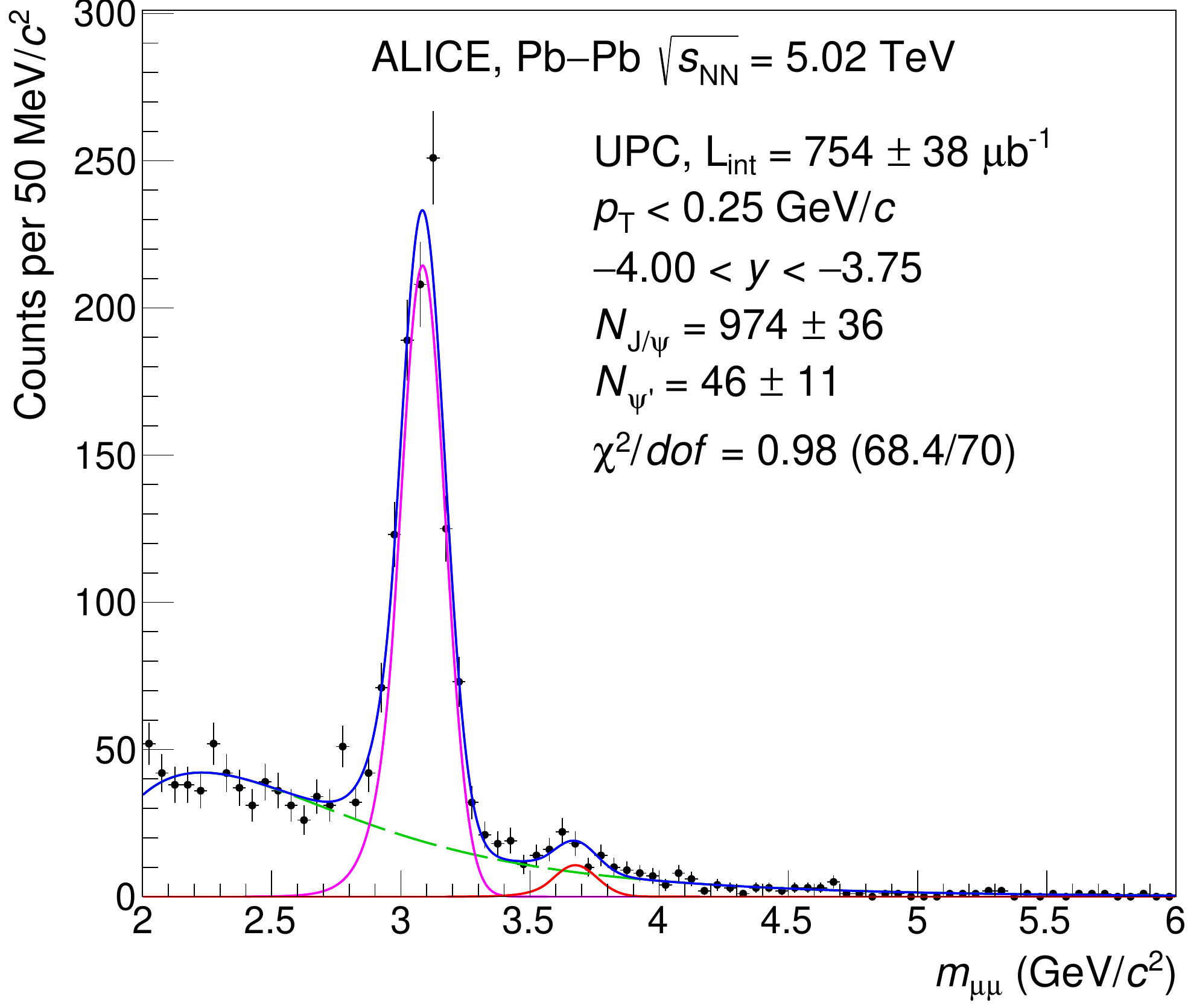} \quad
\includegraphics[width = 0.45\textwidth]{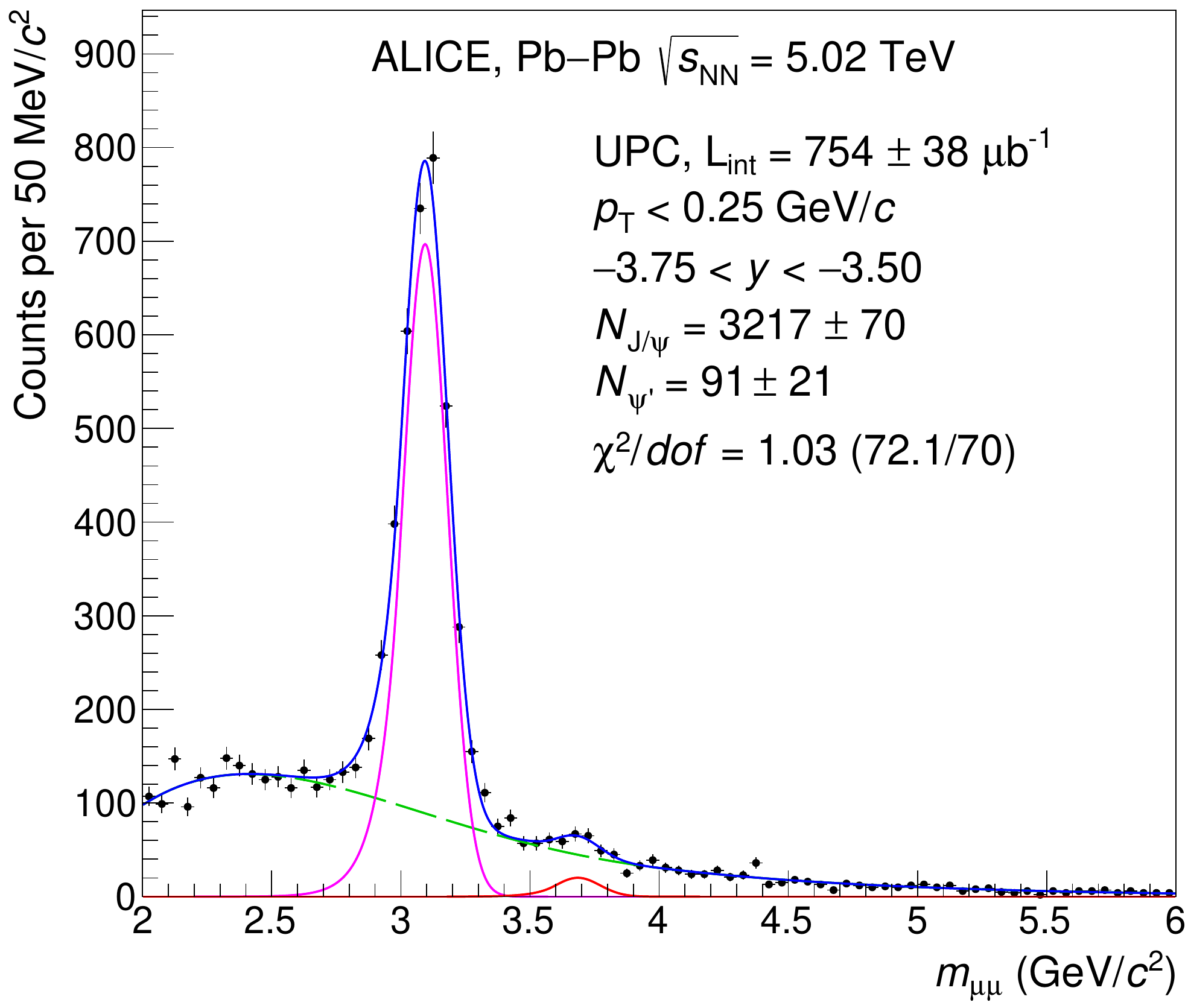}
\includegraphics[width = 0.45\textwidth]{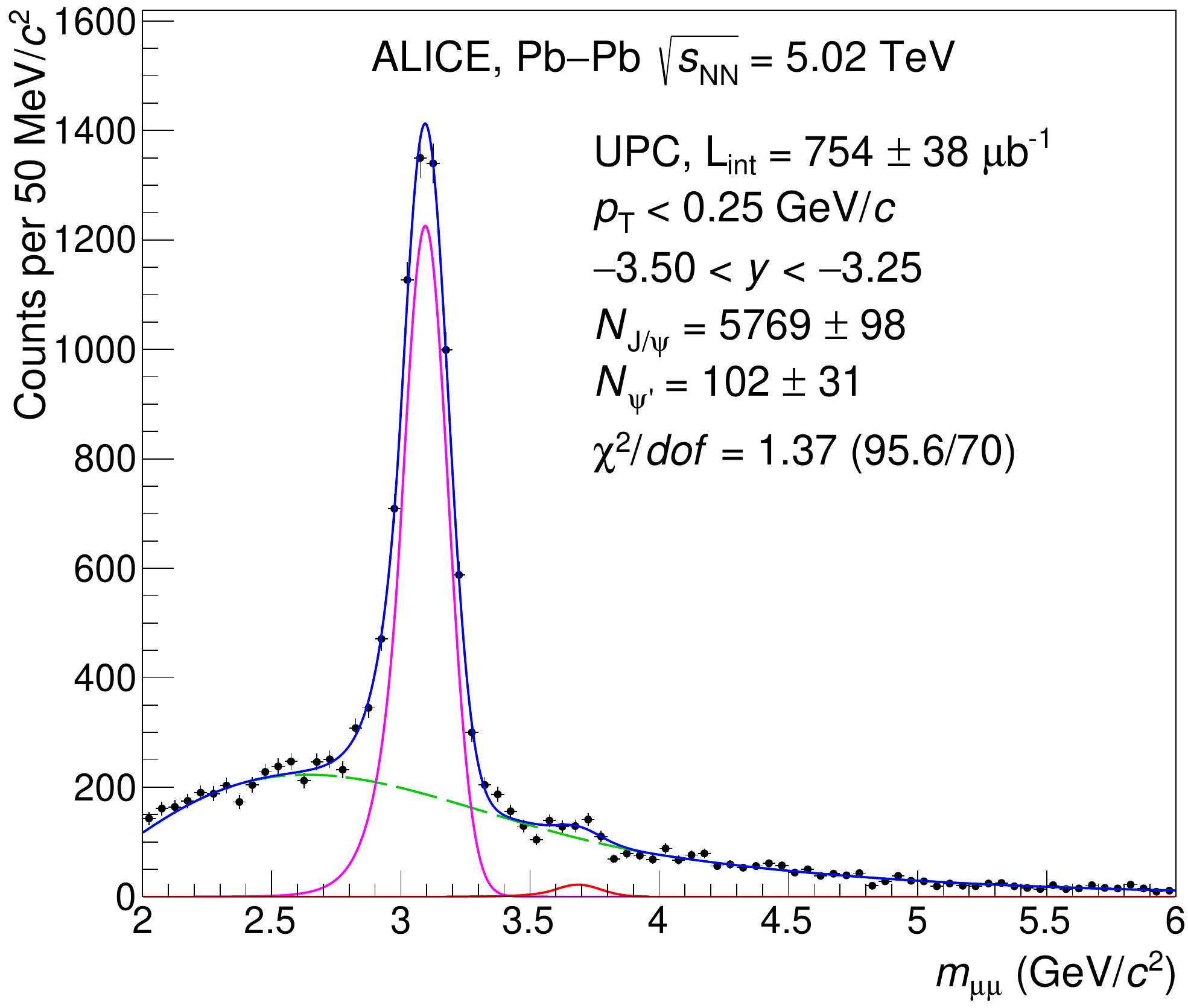}
\quad
\includegraphics[width = 0.45\textwidth]{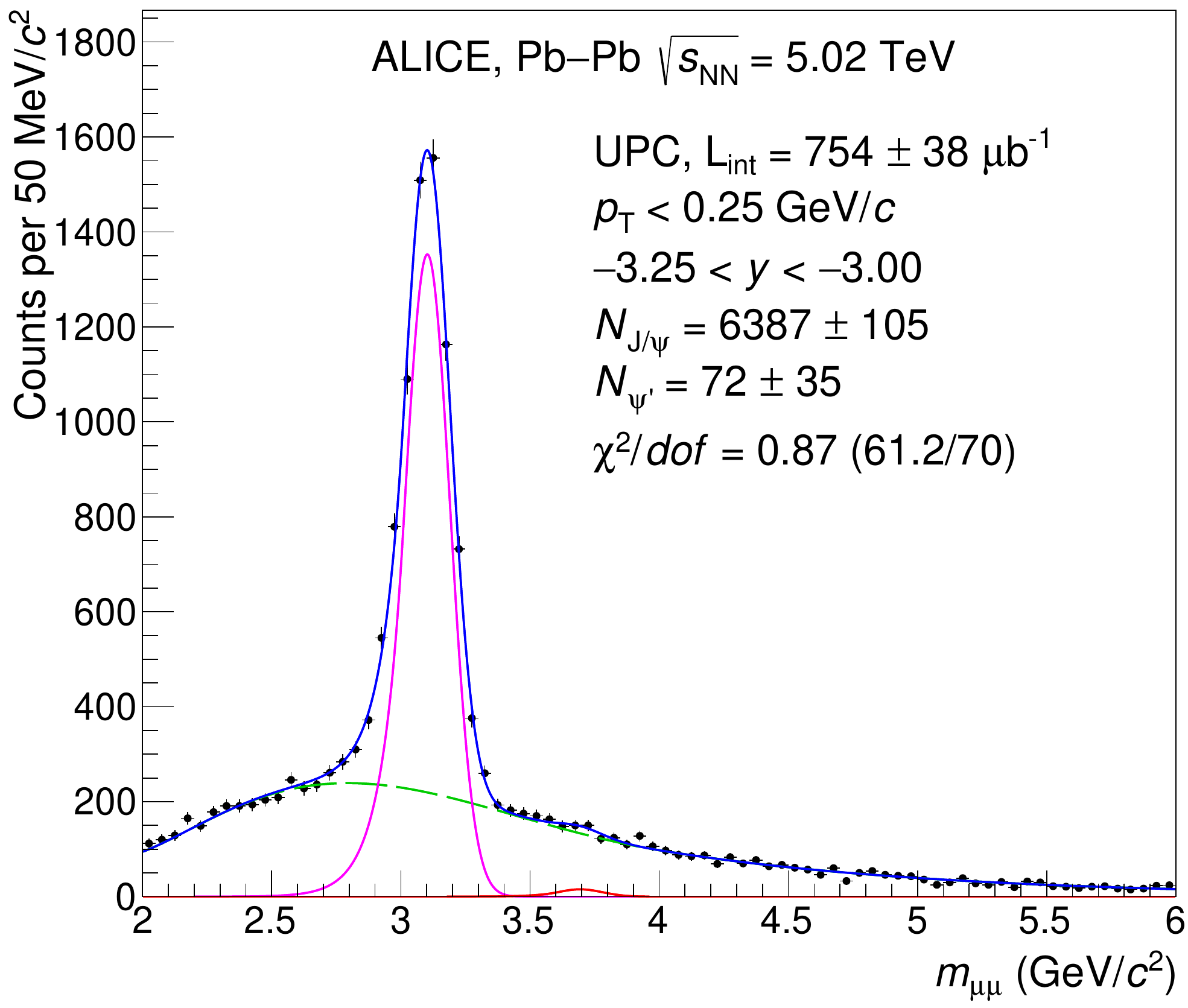}
\includegraphics[width = 0.45\textwidth]{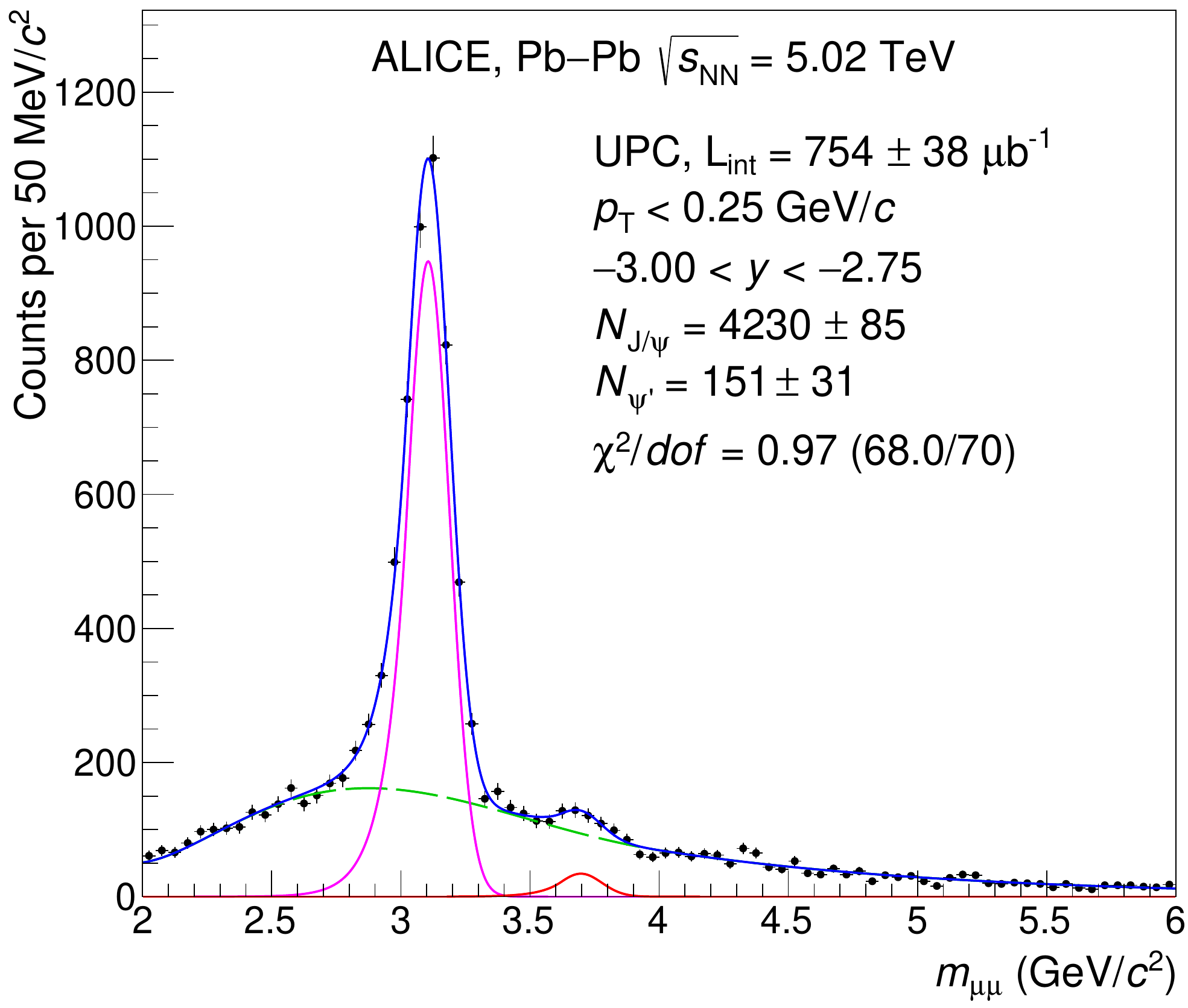}
\quad
\includegraphics[width = 0.45\textwidth]{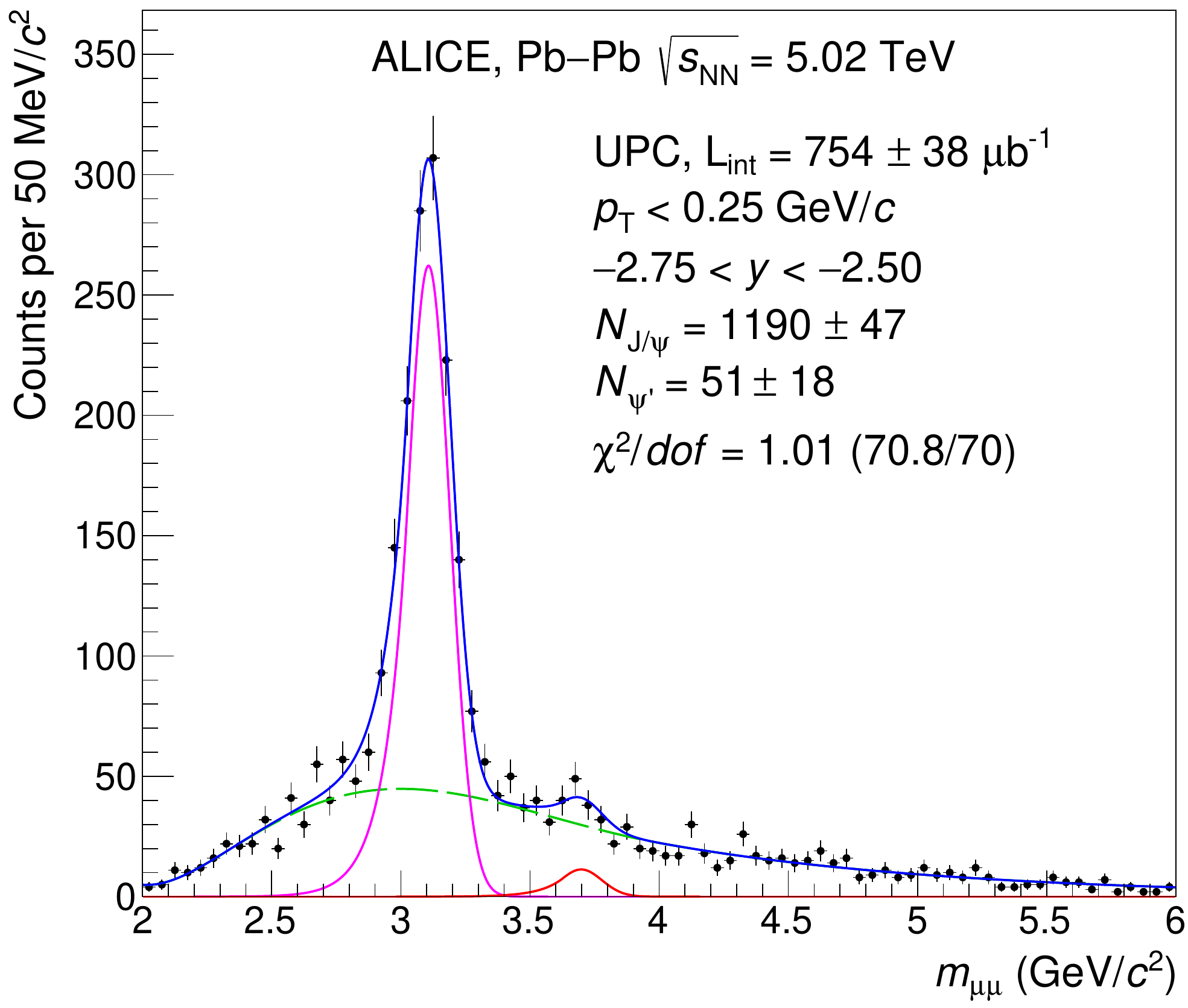}
\end{center}
\caption{Invariant mass distributions in six rapidity bins for muon pairs satisfying the event selection described in the text.}
\label{fig:m}
\end{figure}

A sample enriched in coherent candidates was obtained by selecting dimuons with transverse momentum $\pt<0.25$ \GeVc. The invariant mass distributions for selected unlike-sign muon pairs are shown in Fig.~\ref{fig:m0pt0}, left, in the full dimuon rapidity range $-4.0<y<-2.5$ and in Fig.~\ref{fig:m} in six rapidity subranges. The invariant mass distributions are fitted with a function modeling the background and two Crystal Ball functions~\cite{Gaiser:1982yw} for the \jpsi\ and the \psip peaks. 
The shape of the background  at large invariant masses is well described by an exponential distribution, as expected if it is dominated by the process $\gamma \gamma \to \mu^+ \mu^-$. However, at masses below the \jpsi,\ the distribution is strongly influenced by the muon trigger condition. In order to model this, the whole background distribution is fitted using a template made from reconstructed \starlight events corresponding to the $\gamma \gamma \to \mu^+ \mu^-$ process. The results of the fit are parametrized using a fourth-order polynomial, which turns smoothly into an exponential tail as from 4~GeV/$c^{2}$.
The coefficients of the polynomial are then kept fixed in the fit to the experimental data, while the slope of the exponential term and the normalization are left free. The fitted slope is found to agree within 2.5 standard deviations with the value obtained from the generated sample.

The raw inclusive \jpsi and \psip yields, $N(\jpsi)$ and $N(\psip)$, were obtained by fitting the dimuon invariant mass spectrum in the range $2.2 < m_{\mu\mu} < 6$ GeV/$c^{2}$. The slope parameters in the Crystal Ball functions were fixed from fits to the respective Monte Carlo sets. The width parameter $\sigma_{\jpsi}$ was left free for the \jpsi, and was fixed to $\sigma_{\psip} = \sigma_{\jpsi} \cdot (\sigma_{\psip}^{\rm MC}/\sigma_{\jpsi}^{\rm MC})$ for the \psip, where the ratio $\sigma_{\psip}^{\rm MC}/\sigma_{\jpsi}^{\rm MC} \sim 1.09$ of the \psip to the \jpsi\ widths was obtained from the fits to corresponding Monte Carlo sets. The mass parameter of the Crystal Ball function was left unconstrained for the \jpsi. 
Due to the small \psip statistics, the \psip mass was fixed so that the difference with respect to the \jpsi mass is the same as quoted by the PDG~\cite{Tanabashi:2018oca}.
The \jpsi mass $m_{\jpsi} = 3.0993 \pm 0.0009$  GeV/$c^{2}$, obtained from the fit in the full rapidity range $-4.0<y<-2.5$, is in agreement with the PDG value within 3 standard deviations.

\begin{figure}[h!]
\begin{center}
\includegraphics[width = 0.45\textwidth]{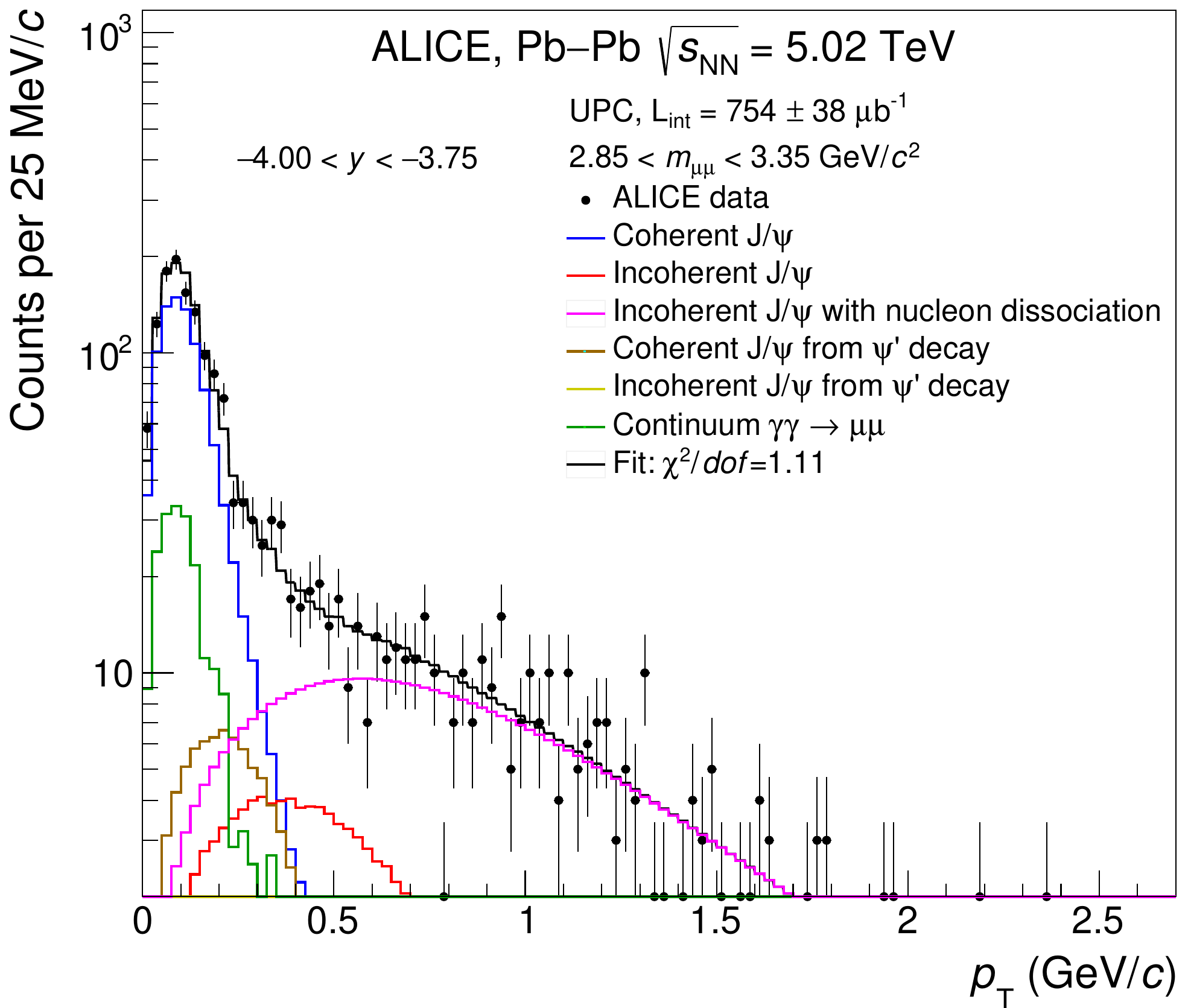}
\quad
\includegraphics[width = 0.45\textwidth]{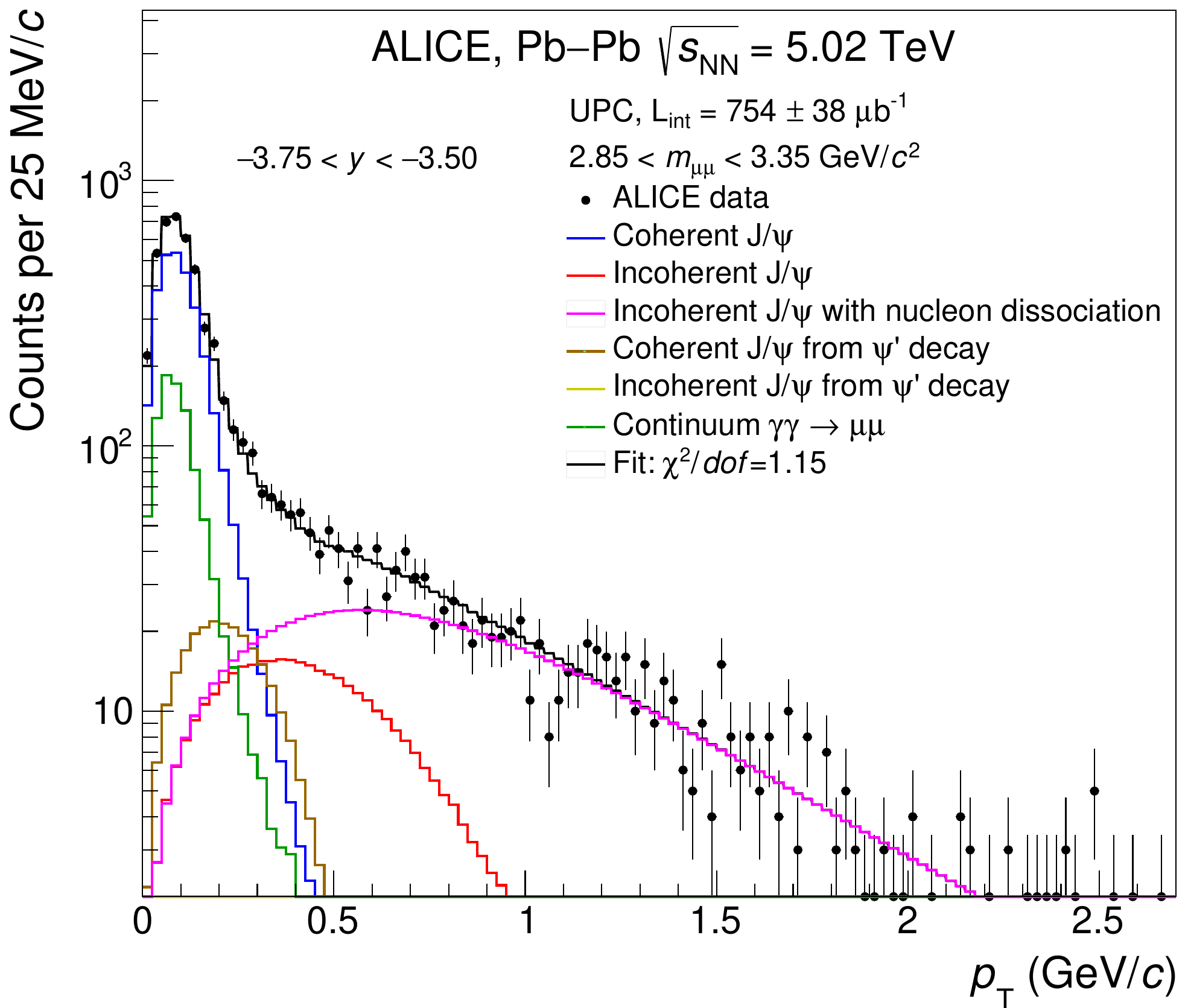}
\includegraphics[width = 0.45\textwidth]{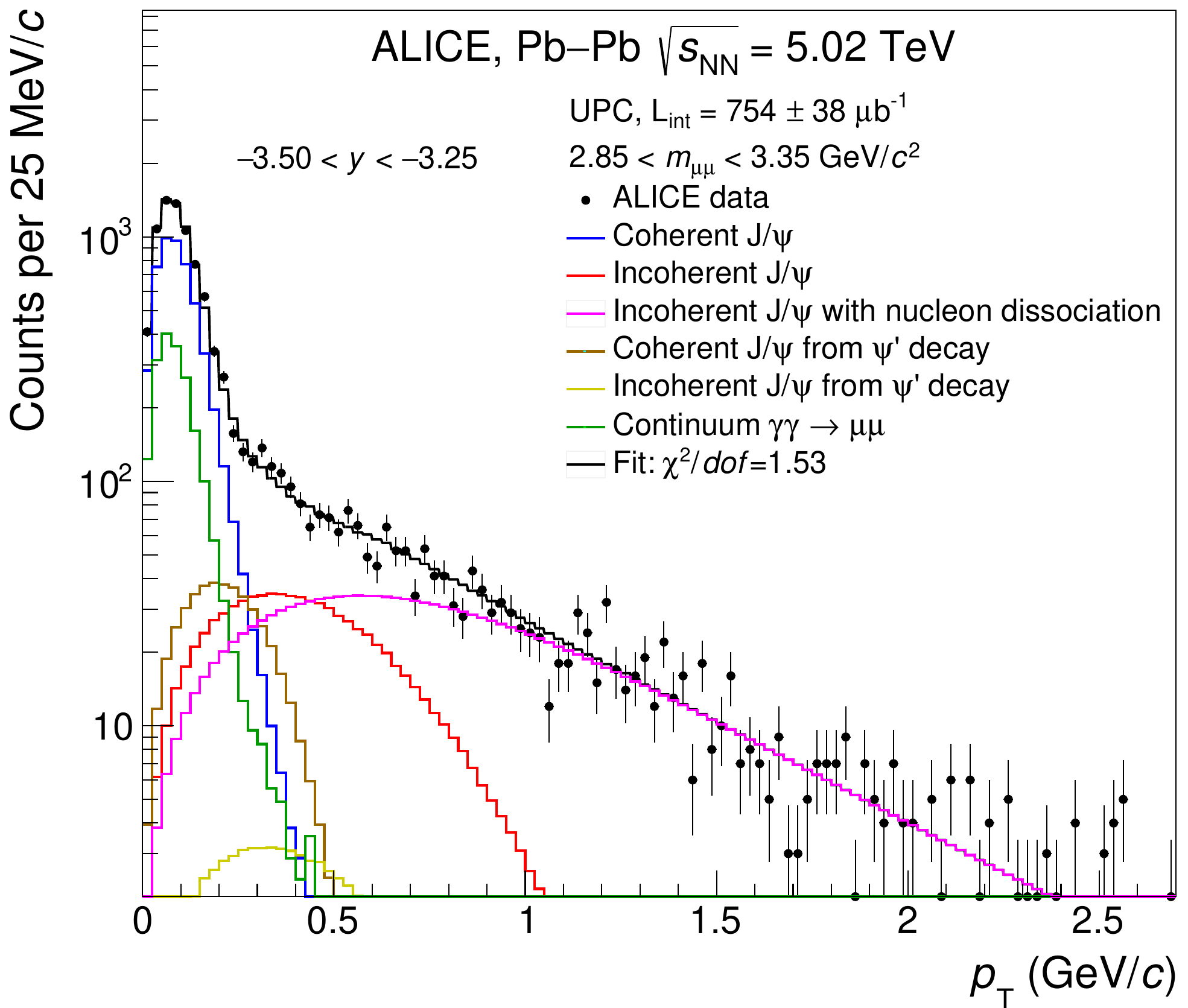}
\quad
\includegraphics[width = 0.45\textwidth]{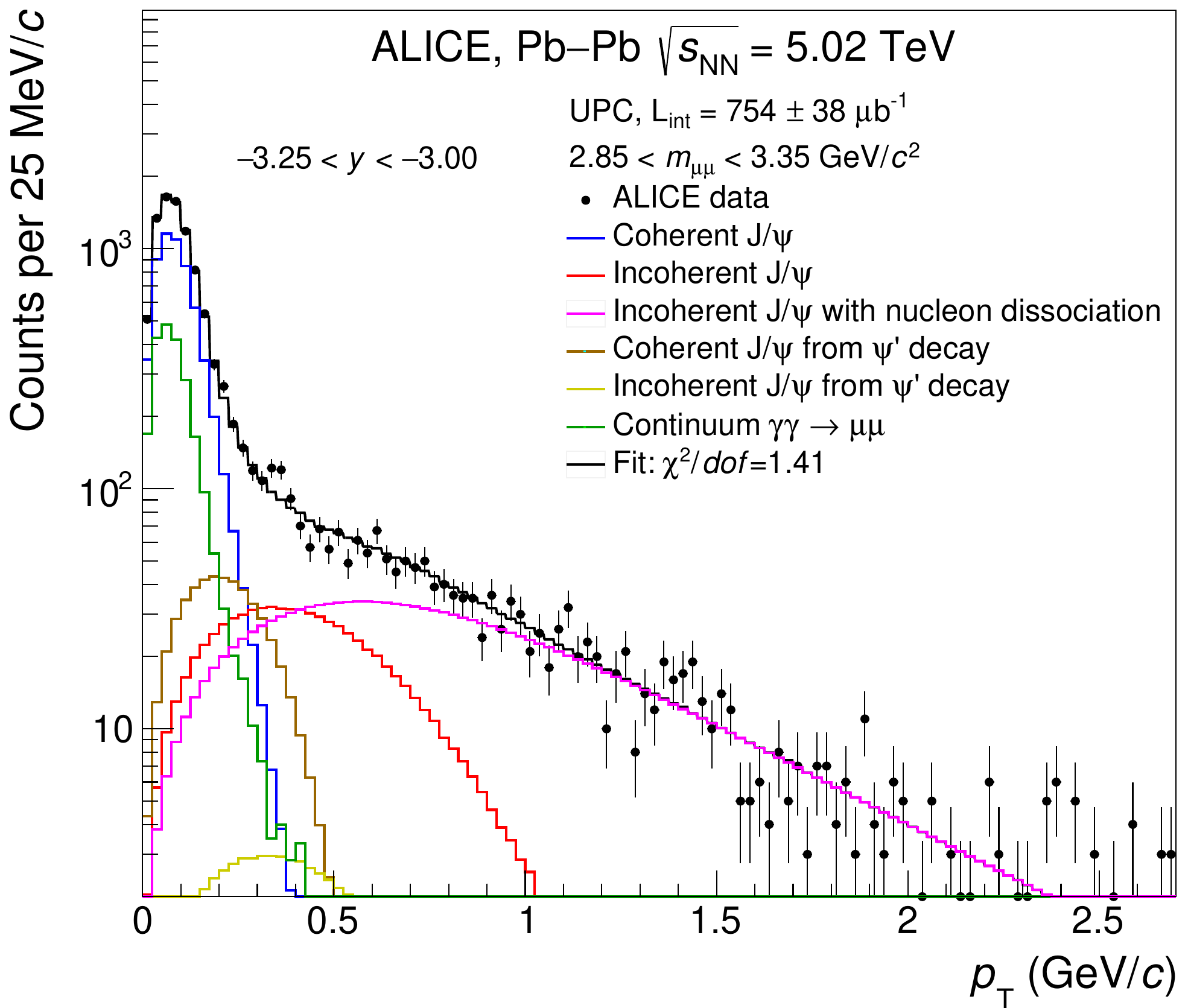}
\includegraphics[width = 0.45\textwidth]{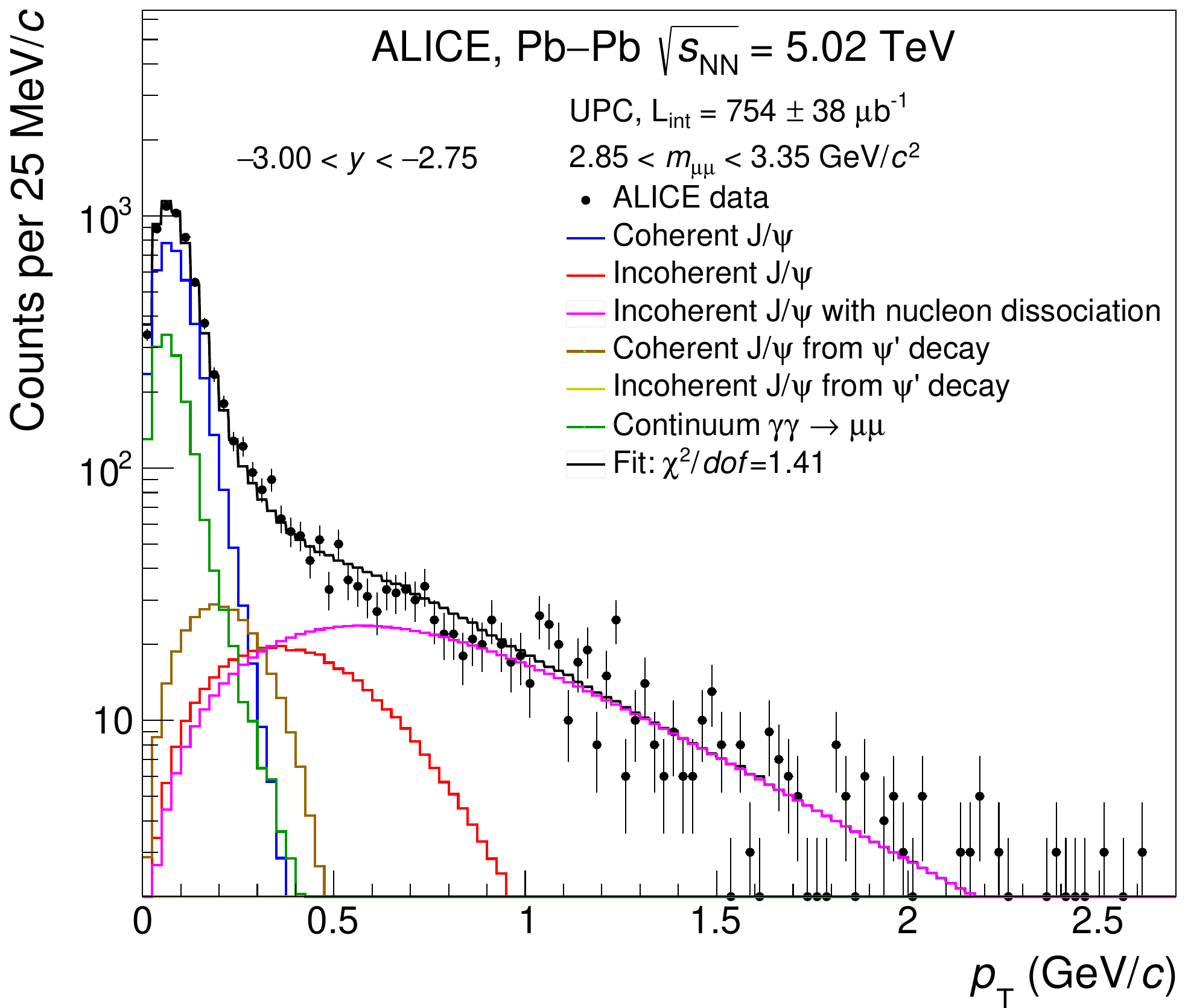}
\quad
\includegraphics[width = 0.45\textwidth]{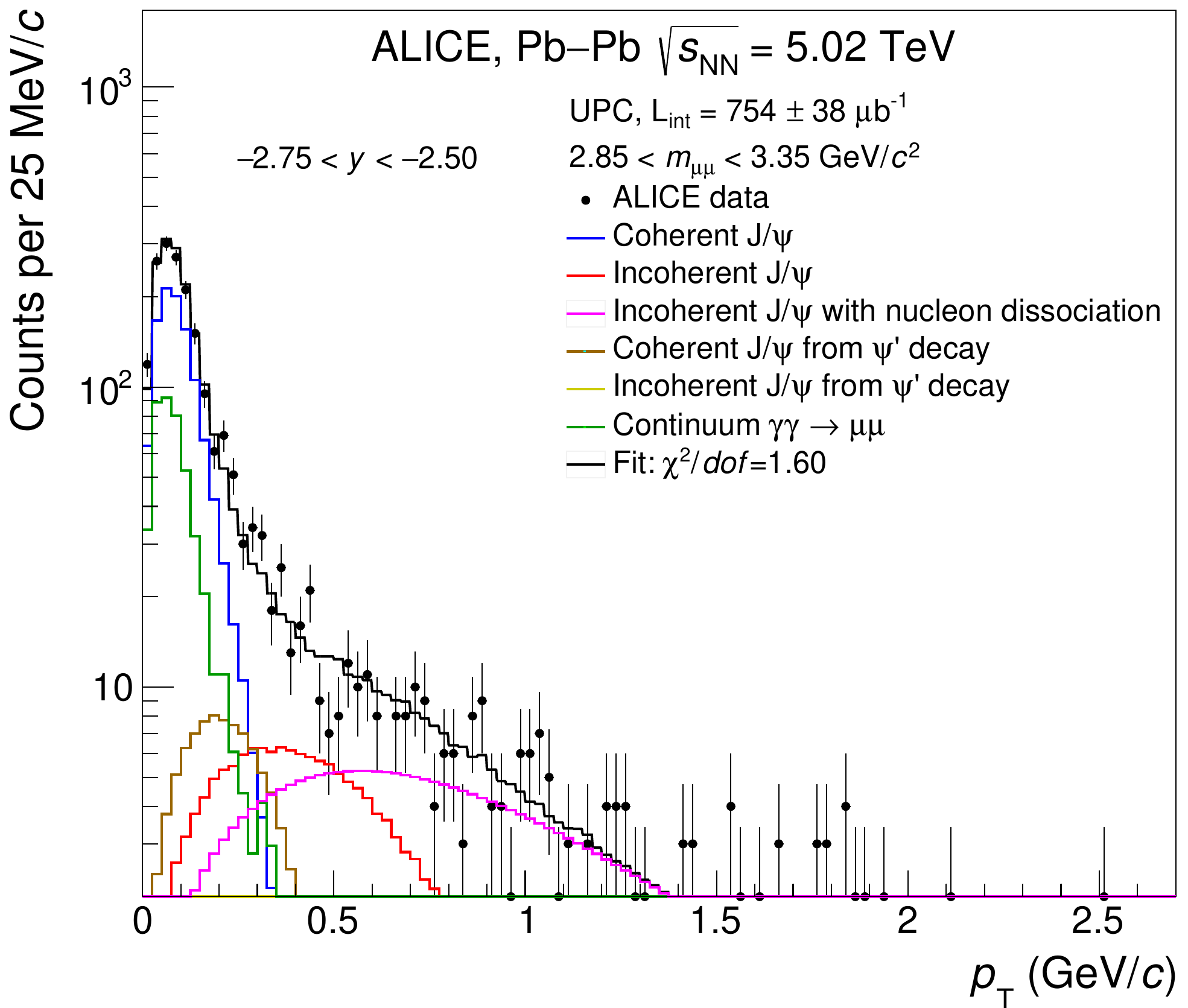}
\end{center}
\caption{
Transverse momentum distributions in six rapidity intervals for muon pairs satisfying the event selection described in the text.}
\label{fig:pt}
\end{figure}

The raw inclusive \jpsi yields obtained from invariant mass fits contain contributions from the coherent and incoherent \jpsi photoproduction, which can be separated in the analysis of transverse momentum spectra. The \pt distributions for dimuons in the range $2.85 < m_{\mu\mu} < 3.35$~GeV/$c^{2}$ are shown in Fig.~\ref{fig:m0pt0}, right, in the full dimuon rapidity range $-4.0<y<-2.5$ and in Fig.~\ref{fig:pt} in six rapidity subranges. These distributions were fitted with Monte Carlo templates produced using  \starlight, corresponding to different production mechanisms: coherent \jpsi, incoherent \jpsi, feed-down \jpsi from coherent \psip decays, feed-down \jpsi from incoherent \psip decays and continuum dimuons from the $\gamma \gamma \to \mu^+ \mu^-$ process. In order to describe the high-$\pt$ tail, the incoherent \jpsi photoproduction accompanied by nucleon dissociation was also taken into account in the fits with the template based on the H1 parametrization of the dissociative \jpsi photoproduction~\cite{Alexa:2013xxa} (denoted as dissociative \jpsi in the following): 
 \begin{equation}
\frac{\mathrm{d}N}{\mathrm{d}\pt} \sim  \pt \left(1 + \frac{b_{\mathrm{pd}}}{n_{\mathrm{pd}}}\pt^{2}\right)^{-n_{\mathrm{pd}}}.
 \end{equation}
\noindent
The H1 collaboration provided two sets of measurements corresponding to different photon--proton center-of-mass energy ranges: $25 {\rm\ \GeV} < W_{\gamma {\rm p}} < 80 {\rm\ \GeV}$ (low-energy data set) and $40 {\rm\ \GeV} < W_{\gamma {\rm p}} < 110 {\rm\ \GeV}$ (high-energy data set). The fit parameters $b_{\rm pd} = 1.79 \pm 0.12$ (GeV$/c)^{-2}$ and $n_{\rm pd}=3.58 \pm 0.15$ from the high-energy data set were used by default, while the corresponding uncertainties and the low-energy values $b_{\rm pd} = 1.6 \pm 0.2$ (GeV$/c)^{-2}$ and $n_{\rm pd}=3.58 ({\rm fixed})$ were used for systematic checks. 

The templates were fitted to the data leaving the normalization free for coherent \jpsi, incoherent \jpsi and dissociative \jpsi production. The normalization of the $\gamma \gamma \to \mu^+ \mu^-$ spectrum was fixed to the one obtained from the invariant mass fits. 
The normalization of the coherent and incoherent feed-down \jpsi templates was constrained to the normalization of primary coherent and incoherent \jpsi templates, according to the feed-down fractions extracted from the measurement of raw inclusive \jpsi and \psip yields, as described below. The extracted incoherent \jpsi  fraction $f_{\rm I} = \frac{N({\rm incoh\ } \jpsi)}{N({\rm coh\ } \jpsi)}$ for $\pt < 0.25$ \GeVc ranges from $(4.9\pm0.6)$\% to $(6.4\pm0.8)$\%  depending on the rapidity interval and is consistent with being constant within the uncertainties of the fits. The contribution of incoherent \jpsi with nucleon dissociation was also taken into account in this fraction.

\section{Results and discussion}

\subsection{Ratio of coherent \psip and \jpsi cross sections}
The obtained dimuon invariant mass spectra can be used to extract the ratio of coherent \psip and \jpsi cross sections $R = \frac{\sigma(\psip)}{\sigma(\jpsi)}$ and the fraction of feed-down \jpsi from \psip decays in the raw \jpsi yields. The fits to the invariant mass distributions for dimuons with pair $\pt<0.25$ \GeVc in the full rapidity range $-4.0<y<-2.5$ result in the following ratio of the measured raw inclusive \psip and \jpsi yields:
\begin{equation}
R_N = \frac{N(\psip)}{N(\jpsi)} = 0.0250 \pm 0.0030 ({\rm stat.}),
\label{eq:R_N_mes}
\end{equation}

The raw \psip and \jpsi yields in this ratio contain contributions both from coherent and incoherent \psip and \jpsi photoproduction. However, according to the dimuon \pt fits, the fraction $\fI$ of incoherent \jpsi in the raw \jpsi yields does not exceed 6\% and, according to \starlight~\cite{Klein:2016yzr} and calculations within the color dipole approach~\cite{Ducati:2013bya}, the fraction of incoherent \psip in the raw \psip yields is expected to be similar. The $R_N$ ratio can therefore be considered as a good estimate of the ratio of coherent \jpsi and \psip yields, since the incoherent fractions of \psip and \jpsi yields largely cancel in the ratio. Besides, the raw \jpsi yields contain significant feed-down contribution coming from $\psip \to \jpsi + {\rm anything}$ decays. Taking into account this feed-down contribution, one can express the $R_N$ ratio in terms of primary coherent \psip and \jpsi photoproduction cross sections $\sigma(\psip)$ and  $\sigma(\jpsi)$ integrated over all transverse momenta in the rapidity range $-4.0<y<-2.5$:
\begin{equation}
R_N 
= \frac{\sigma(\psip) BR(\psip \to \mu\mu)\epsilon(\psip) }{\sigma(\jpsi) BR(\jpsi \to \mu\mu)\epsilon(\jpsi) + \sigma(\psip) BR(\psip \to \jpsi)\epsilon(\psip \to \jpsi) BR(\jpsi \to \mu\mu)}
\label{eq:R_N}
\end{equation}
where $\epsilon(\jpsi) = 12.0\%$, $\epsilon(\psip) = 15.8\%$ and $\epsilon(\psip \to \jpsi) = 7.2\%$ are the efficiency corrections for primary coherent \jpsi, \psip and feed-down $\jpsi$ from coherent \psip decays estimated with \starlight, while $BR(\jpsi\to \mu\mu) = (5.961 \pm 0.033)\%$, $BR(\psip \to \mu\mu) = (0.80 \pm 0.06)\%$, $BR(\psip \to \jpsi + {\rm anything}) = (61.4 \pm 0.6)\%$ are the corresponding branching ratios~\cite{Tanabashi:2018oca}. 
Equation~(\ref{eq:R_N}) can be used to express the ratio of primary coherent \psip and \jpsi photoproduction cross sections, $R$, in terms of the measured yield ratio $R_N$:
\begin{equation}
R = 
\frac{R_N BR(\jpsi \to \mu\mu)\epsilon(\jpsi)}
{
BR(\psip \to \mu\mu)\epsilon(\psip)-R_N BR(\psip \to \jpsi)\epsilon(\psip \to \jpsi) BR(\jpsi \to \mu\mu)
}
\label{eq:R}
\end{equation}
Substituting the measured $R_N$ value from Eq.~(\ref{eq:R_N_mes}) and the corresponding efficiency values and branching ratios, one gets:
\begin{equation}
R = 0.150 \pm 0.018 ({\rm stat.}) \pm 0.021 ({\rm syst.}) \pm 0.007 ({\rm BR}),
\end{equation}
where the uncertainties on branching ratios $BR(\jpsi\to \mu\mu)$ and $BR(\psip \to \mu\mu)$ were added in quadrature, while the main sources of systematic uncertainties are the variation of the fit range, of the signal and background shapes, and of the dimuon transverse momentum cut.

The measured ratio of the \psip and \jpsi cross sections is compatible with the exclusive photoproduction cross section ratio $R = 0.166 \pm 0.007({\rm stat.}) \pm 0.008({\rm syst.}) \pm 0.007({\rm BR})$ measured by the H1 collaboration in ep collisions~\cite{Adloff:2002re} and with the ratio $R\approx0.19$ measured by the LHCb collaboration  in \pp collisions~\cite{Aaij:2018arx}. The measured ratio also agrees with predictions based on the Leading Twist Approximation~\cite{Guzey:2016piu} for \PbPb UPC ranging from 0.13 to 0.18 depending on the model parameters.
The \psip-to-\jpsi coherent cross section ratio is expected to have a mild  dependence on the collision energy and vector meson rapidity~\cite{Guzey:2016piu} (at most a few percent). Therefore the measured ratio can be directly compared to the unexpectedly large \psip-to-\jpsi coherent cross section ratio $0.34^{+0.08}_{-0.07}$, measured by ALICE in the $\psi' \to l^+l^-$ and $\psi' \to l^+l^- \pi^+\pi^-$ channels at central rapidity in \PbPb UPC at \twosevensixnn~\cite{Adam:2015sia}. The ratio at central rapidity is more than a factor two larger but still stays compatible within 2.5 standard deviations with the forward rapidity measurement, owing mainly to the large uncertainties of the central rapidity measurement that will be improved by the analysis of the much larger UPC data sample collected with the ALICE central barrel in Run 2.

The measured cross section ratio $R$ was used to extract the fraction of feed-down \jpsi from \psip relative to the primary \jpsi yield:
\begin{equation}
f_{\rm D} = \frac{N (\mbox{feed-down\ } \jpsi)}{N (\mbox{\rm primary\ } \jpsi)} = 
R \frac{\epsilon(\psip \to \jpsi)}{\epsilon(\jpsi)} BR(\psip \to \jpsi) 
\end{equation}
The fraction $f_{\rm D} = 8.5\%\pm1.5\%$ was obtained for the full rapidity range without any $\pt$ cut, where statistical, systematic and branching ratio uncertainties were added in quadrature. The fraction reduces to $f_{\rm D} = 5.5\% \pm 1.0\%$ for $\pt <0.25$ \GeVc because feed-down \jpsi are characterized by wider transverse momentum distributions compared to primary \jpsi.

\subsection{Coherent \jpsi cross section}

The coherent \jpsi differential cross section is given by:
\begin{equation}
\frac{\rm{d}{\sigma^{\rm coh}_{\jpsi}}}{\rm{d}y} = \frac{N(\jpsi)}{(1+f_{\rm I} + f_{\rm D})\epsilon(\jpsi) \rm{BR}(\jpsi \to \mu\mu) \epsilon_{\rm veto} L_{\rm int} \Delta y}
\label{eq_cs}
\end{equation}
The  raw $\jpsi$ yield values, efficiencies, $f_{\rm I}$ and $f_{\rm D}$ fractions and coherent $\jpsi$ cross sections with relevant statistical and systematic uncertainties are summarized in~Table~\ref{tab:cs}.  The associated systematic uncertainties are briefly described in the following.

\begin{table}
\centering
\caption{$\jpsi$ yields, efficiencies, $f_{\rm I}$ and $f_{\rm D}$ fractions and coherent $\jpsi$ cross sections.}
\begin{tabular}{cccccc}
\hline
rapidity range  &  $N_{\jpsi}$     & $\epsilon$& $\fd$    & $\fI$        & ${\rm d}\sigma^{\rm coh}_{\jpsi}/{\rm d}y$ (mb) \\
\hline
 $(-4.00, -2.50)$&  $21747 \pm 190 $& $0.120 $ & $0.055 $ & $0.055 \pm 0.001$ & $2.549 \pm 0.022\ ({\rm stat.}) {\ }_{-0.237}^{+0.209}\ ({\rm syst.})$\\
 $(-4.00, -3.75)$&  $  974 \pm  36 $& $0.051 $ & $0.055 $ & $0.064 \pm 0.008$ & $1.615 \pm 0.060\ ({\rm stat.}) {\ }_{-0.147}^{+0.135}\ ({\rm syst.})$\\
 $(-3.75, -3.50)$&  $ 3217 \pm  70 $& $0.140 $ & $0.055 $ & $0.058 \pm 0.004$ & $1.938 \pm 0.042\ ({\rm stat.}) {\ }_{-0.190}^{+0.166}\ ({\rm syst.})$\\
 $(-3.50, -3.25)$&  $ 5769 \pm  98 $& $0.204 $ & $0.055 $ & $0.060 \pm 0.003$ & $2.377 \pm 0.040\ ({\rm stat.}) {\ }_{-0.229}^{+0.212}\ ({\rm syst.})$\\
 $(-3.25, -3.00)$&  $ 6387 \pm 105 $& $0.191 $ & $0.055 $ & $0.052 \pm 0.002$ & $2.831 \pm 0.047\ ({\rm stat.}) {\ }_{-0.280}^{+0.253}\ ({\rm syst.})$\\
 $(-3.00, -2.75)$&  $ 4229 \pm  85 $& $0.119 $ & $0.055 $ & $0.049 \pm 0.003$ & $3.018 \pm 0.061\ ({\rm stat.}) {\ }_{-0.294}^{+0.259}\ ({\rm syst.})$\\
 $(-2.75, -2.50)$&  $ 1190 \pm  47 $& $0.029 $ & $0.054 $ & $0.049 \pm 0.006$ & $3.531 \pm 0.139\ ({\rm stat.}) {\ }_{-0.362}^{+0.294}\ ({\rm syst.})$\\
 \hline
\end{tabular}
\label{tab:cs}
\end{table}
The first source of systematic uncertainty is related to the separation of peripheral and ultra-peripheral collisions. Coherent-like $\jpsi$ photoproduction, observed in peripheral collisions of heavy ions~\cite{Adam:2015gba}, may contribute a few per cent to the raw $\jpsi$ yields in case hadronic activity is not detected by the \VZERO and \AD detectors. In order to reduce a possible contamination from $\jpsi$ produced in peripheral hadronic events, the analysis was repeated with an additional requirement that there be no tracklets detected at mid-rapidity in the SPD (where a tracklet is a segment formed by at least one hit in each of the two detector layers), resulting in 12.6\% to 15.0\% lower \jpsi yields depending on the rapidity range. The veto inefficiency associated with this additional SPD requirement was estimated with unbiased triggers similar to what was done for the \VZERO and \AD veto inefficiencies. The average fraction of events with at least one SPD tracklet was found to be $p_{\rm SPD} = 9.4 \pm 0.2\%$. The yields corrected for the additional SPD veto inefficiency of 9.4\% result in cross sections 3.6\% to 6.0\% lower than the ones obtained without the SPD veto. This cross section difference is taken into account in the systematic uncertainty.

The systematic uncertainties on the efficiencies obtained by variation
of the generated rapidity shapes range from 0.1\% to 0.8\%, depending on the rapidity interval. The tracking efficiency uncertainty of 3\% was estimated by comparing the single-muon tracking efficiency values obtained in MC and data, with a procedure that exploits the redundancy of the tracking-chamber information~\cite{Adam:2015isa}. The systematic uncertainty on the dimuon trigger efficiency has two origins: the intrinsic efficiencies of the muon trigger chambers and the response of the trigger algorithm. 
The first one was determined by varying the trigger chamber efficiencies in the MC by an amount equal to the statistical uncertainty on their measurement with a data-driven method and amounts to 1.5\%.  The second one was estimated by comparing the trigger response function between data and MC, resulting in efficiency differences ranging from 5\% to 6\% depending on the rapidity interval. Finally, there is a 1\% contribution related to the precision required to match track segments reconstructed in the tracking and trigger chambers.

Several tests were performed to estimate the uncertainty on the raw \jpsi signal extraction. These include the uncertainty on the $\jpsi$ signal shape estimated by fitting the Crystal Ball slope parameters instead of fixing them from Monte-Carlo templates and by replacing the single-sided Crystal Ball with a double-sided Crystal Ball function. The variation of the continuum background shape due to the uncertainty on the trigger response function, variation of the invariant mass intervals by $\pm 0.1$ GeV/$c^{2}$ and of the dimuon \pt selection by $\pm 0.05$ GeV/$c$ were also considered. The systematic uncertainty on the raw \jpsi yield, estimated as root mean square of the results obtained from all tests, is about 2\% with a slight rapidity dependence.

Several sources of systematic uncertainties are associated with different contributions to the \pt spectrum: the fraction of feed-down \jpsi, the shape and contribution of the $\gamma \gamma \to \mu^+ \mu^-$ template, the shape for the coherent \jpsi and the shape for the incoherent \jpsi with nucleon dissociation. These contributions are shortly detailed in the following. First, the fraction $\fd$ of feed-down \jpsi with \pt below 0.25 \GeVc was varied in the range from 4.4 to 6.4\% corresponding to the total systematic uncertainty of the measured \psip-to-\jpsi cross section ratio. Second, the shape of the $\gamma \gamma \to \mu^+ \mu^-$ \pt template from \starlight does not include possible contributions from incoherent emission of photons, characterized by  much wider transverse momentum distributions extending well above 1 \GeVc. In order to account for these contributions, 
the shape of the $\gamma \gamma \to \mu^+ \mu^-$ \pt template was changed from \starlight to that obtained from the side-bands surrounding the \jpsi\ peak in the invariant mass spectra, resulting in 1\% systematic uncertainty on the measured coherent cross section. Third, a 0.2\% systematic uncertainty was determined via the variation of the $\gamma\gamma$ contribution according to the statistical uncertainty in the background term calculated from the invariant mass fits. A modification of the transverse momentum spectra for the coherent \jpsi according to the model~\cite{Guzey:2016qwo}, results in a 0.1\% systematic uncertainty.
Finally, the template shape for the incoherent \jpsi with nucleon dissociation was varied by exchanging the H1 high-energy run parameters for those determined from the low-energy run resulting in a 0.1\% systematic uncertainty on the coherent cross section. 

The systematic uncertainties are summarized in Table~\ref{tab:syst}. The total systematic uncertainty is the quadratic sum of all the sources listed in the table. Luminosity normalization, veto efficiency and branching ratio uncertainties are fully correlated. The uncertainty on the signal extraction is considered as uncorrelated as a function of rapidity. Finally, all other sources of uncertainty are considered as partially correlated across  different rapidity intervals. 

\begin{table}[t]
\centering
\caption{Summary of systematic uncertainties. The ranges of values correspond to different rapidity bins.}
\begin{tabular}{|l|c|}
\hline
Source               & Value  \\
\hline
Lumi. normalization  & $\pm 5.0\%$ \\
Branching ratio      & $\pm0.6\%$\\
SPD, V0 and AD veto  & from $-3.6$\% to $-6.0$\% \\
MC rapidity shape    & from $\pm0.1$\% to $\pm0.8$\% \\
Tracking             & $\pm3.0\%$\\
Trigger              & from $\pm 5.2\%$ to $\pm 6.2\%$\\
Matching             & $\pm1.0\%$\\
Signal extraction    & $\pm2.0\%$\\
$\fd$ fraction       & $\pm0.7\%$\\
$\gamma\gamma$ yield & $\pm1.2\%$\\
$p_{\rm T}$ shape for coherent $\jpsi$  & $\pm0.1\%$\\
$b_{\rm pd}$ parameter & $\pm0.1\%$\\
\hline            
Total                & from ${}^{+8.3}_{-9.2}\%$ to  ${}^{+8.9}_{-10.3}\%$  \\
\hline
\end{tabular}
\label{tab:syst}
\end{table}

\subsection{Discussion}

\begin{figure}[tb]
\begin{center}
\includegraphics[width = 0.70\textwidth]{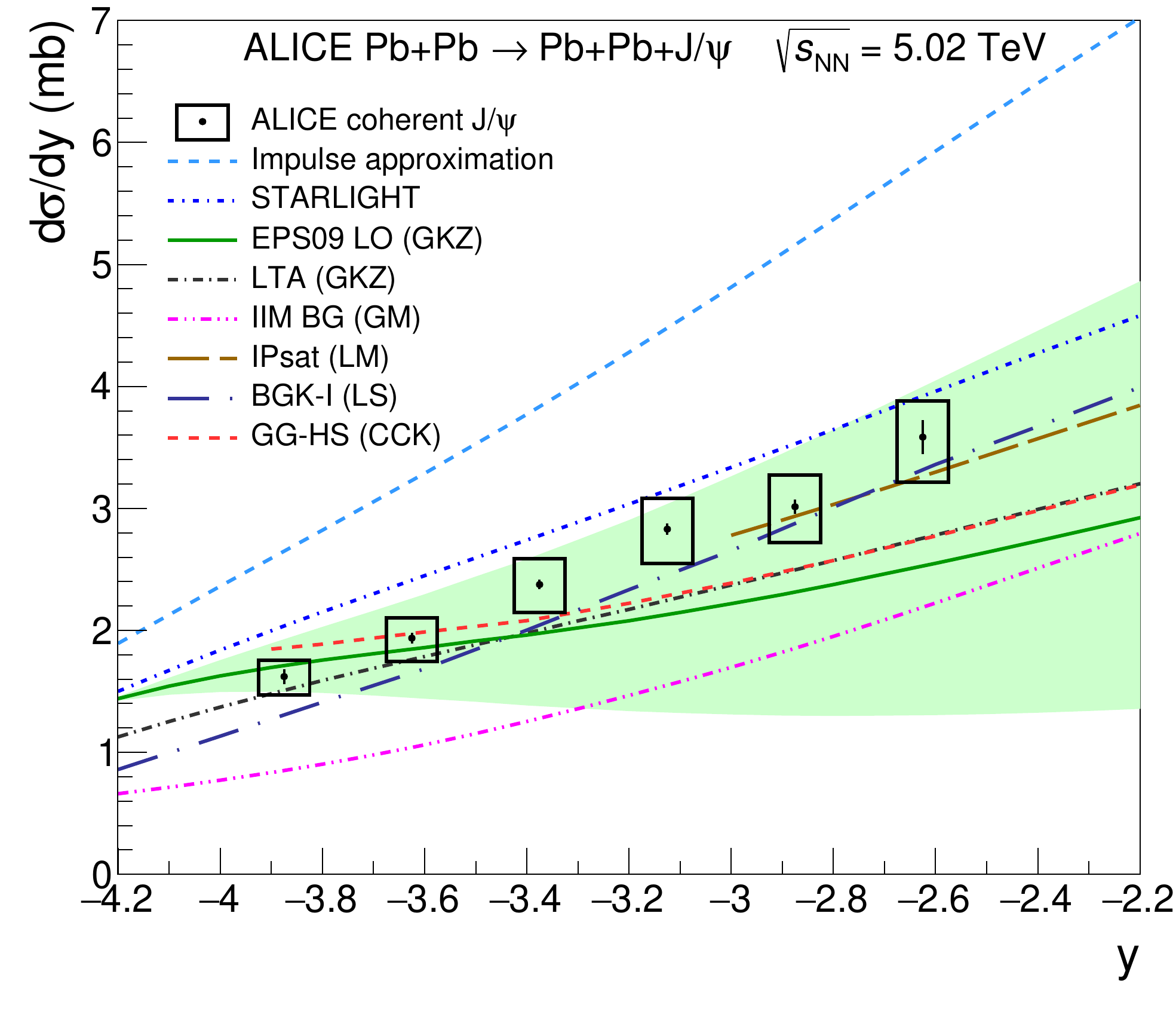}
\end{center}
\caption{Measured coherent differential cross section of \jpsi~photoproduction in ultra-peripheral \PbPb collisions at \fivenn. The error bars represent the statistical uncertainties, the boxes around the points the systematic uncertainties. The theoretical calculations~\cite{Klein:1999qj,Guzey:2013xba,Klein:2016yzr,Guzey:2016piu,Goncalves:2014wna,Santos:2014zna,Lappi:2010dd,Lappi:2013am,Cepila:2016uku,Cepila:2017nef} described in the text are also shown. The green band represents the uncertainties of the EPS09 LO calculation.}
\label{fig:cs}
\end{figure}

The measured differential cross section of coherent \jpsi photoproduction in the rapidity range $-4.0 < y <-2.5$ is shown in Fig.~\ref{fig:cs} and compared with various models. The covered rapidity range corresponds to a Bjorken-$x$ of gluons either in the range $1.1 \cdot 10^{-5}< x < 5.1\cdot 10^{-5}$ or $0.7\cdot 10^{-2} < x < 3.3 \cdot 10^{-2}$ depending on which nucleus emitted the photon. According to models~\cite{Guzey:2016piu}, the fraction of high Bjorken-$x$ gluons ($x \sim 10^{-2}$) is dominant at forward rapidities and ranges from  ${\sim}60\%$ at $y=-2.5$ to ${\sim}95\%$ at $y=-4$.

The Impulse Approximation, taken from \starlight~\cite{Klein:1999qj}, is based on the data from the exclusive \jpsi photoproduction off protons and neglects all nuclear effects except for coherence. The square root of the ratio of experimental points and the Impulse Approximation cross section is about 0.8, reflecting the magnitude of the nuclear gluon shadowing factor at typical Bjorken-$x$ values around $10^{-2}$, under the assumption that the contribution from low Bjorken-$x \sim 10^{-5}$ can be neglected~\cite{Guzey:2013xba}.

\starlight is based on the Vector Meson Dominance model and a parametrization of the existing data on  \jpsi photoproduction off protons~\cite{Klein:2016yzr}. A Glauber-like formalism is used to calculate the \jpsi photoproduction cross section in \PbPb UPC accounting for multiple interactions within the nucleus but not accounting for gluon shadowing corrections. The \starlight model overpredicts the data, indicating the importance of gluon shadowing effects, but the discrepancy is much lower than for the Impulse Approximation.

Guzey, Kryshen and Zhalov \cite{Guzey:2016piu} provide two calculations (GKZ), one based on the EPS09 LO parametrization of the available nuclear shadowing data~\cite{Eskola:2009uj} and the other on the Leading Twist Approximation (LTA) of nuclear shadowing based on the combination of the Gribov-Glauber theory and the diffractive PDFs from HERA~\cite{Frankfurt:2011cs}. Both the LTA model and the EPS09 curve, corresponding to the EPS09 LO central set, underpredict the data but remain compatible with it at the most forward rapidities. The data tends to follow the upper limit of uncertainties of the EPS09 calculation corresponding to the upper bound of uncertainties on the gluon shadowing factor in the EPS09 LO framework. 

Several theoretical groups provided predictions within the color dipole approach coupled to the Color Glass Condensate (CGC) formalism with different assumptions on the dipole-proton scattering amplitude. Predictions by Gon\c{c}alves, Machado et al. (GM) based on IIM and b-CGC models for the scattering amplitude underpredict the data~\cite{Goncalves:2014wna,Santos:2014zna}. Predictions by Lappi and M\"{a}ntysaari (LM) based on the IPsat model~\cite{Lappi:2010dd,Lappi:2013am} give reasonable agreement though the range of predictions does not span all the experimental points. Recent predictions by Luszczak and Schafer (LS BGK-I) within the color-dipole formulation of the Glauber-Gribov theory~\cite{Luszczak:2019vdc} are in agreement with data at semi-forward rapidities, $|y|<3$, but slightly underpredict the data at more forward rapidities.

Cepila, Contreras and Krelina (CCK) provided two predictions based on the extension of the energy-dependent hot-spot model~\cite{Cepila:2016uku} to the nuclear case: using the standard Glauber-Gribov formalism (GG-HS) and using geometric scaling (GS-HS) to obtain the nuclear saturation scale~\cite{Cepila:2017nef}. The GG-HS model agrees with data at most forward rapidities but underpredicts it at semi-forward rapidities. The GS-HS model (not shown) strongly underpredicts the data.

\section{Conclusions}

The first rapidity-differential measurement on the coherent photoproduction of $\jpsi$ in the rapidity interval $-4<y<-2.5$ in ultra-peripheral \PbPb collisions at \fivenn has been presented and compared with model calculations. 
The Impulse Approximation and \starlight models overpredict the data, indicating the importance of gluon shadowing effects. The model based on the central set  of the EPS09 gluon shadowing parametrization, the Leading Twist Approximation, and the hot-spot model coupled to the Glauber-Gribov formalism underpredict the data but remain compatible with it at most forward rapidities. The majority of color dipole models underpredict the data. 

The nuclear gluon shadowing factor of about 0.8 at Bjorken-$x$ values around $10^{-2}$ and a hard scale around the charm quark mass was estimated from the comparison of the measured coherent \jpsi cross section with the Impulse Approximation under the assumption that the contribution from low Bjorken $x \sim 10^{-5}$ can be neglected. Future studies on coherent heavy vector meson photoproduction accompanied by neutron emission may help to decouple low-$x$ and high-$x$ contributions and provide valuable constraints on poorly known gluon shadowing effects at Bjorken $x \sim 10^{-5}$~\cite{Citron:2018lsq}. 

The ratio of the \psip and \jpsi cross sections is in reasonable agreement both with the ratio of photoproduction cross sections off protons measured by the H1 and LHCb collaborations and with LTA predictions for \PbPb UPC.


\newenvironment{acknowledgement}{\relax}{\relax}
\begin{acknowledgement}
\section*{Acknowledgements}
\input{fa_2019-03-29.tex}
\end{acknowledgement}

\bibliographystyle{utphys}   
\bibliography{bibliography}

\newpage
\appendix

%
%

\section{The ALICE Collaboration}
\label{app:collab}
\input{2019-03-29-Alice_Authorlist_2019-Mar-29.tex}  
\end{document}

%% file: commands.tex
%

\newcommand{\pp}           {pp\xspace}
\newcommand{\ppbar}        {\mbox{$\mathrm {p\overline{p}}$}\xspace}
\newcommand{\XeXe}         {\mbox{Xe--Xe}\xspace}
\newcommand{\PbPb}         {\mbox{Pb--Pb}\xspace}
\newcommand{\pA}           {\mbox{pA}\xspace}
\newcommand{\pPb}          {\mbox{p--Pb}\xspace}
\newcommand{\AuAu}         {\mbox{Au--Au}\xspace}
\newcommand{\dAu}          {\mbox{d--Au}\xspace}

\newcommand{\s}            {\ensuremath{\sqrt{s}}\xspace}
\newcommand{\snn}          {\ensuremath{\sqrt{s_{\mathrm{NN}}}}\xspace}
\newcommand{\pt}           {\ensuremath{p_{\rm T}}\xspace}
\newcommand{\meanpt}       {$\langle p_{\mathrm{T}}\rangle$\xspace}
\newcommand{\ycms}         {\ensuremath{y_{\rm CMS}}\xspace}
\newcommand{\ylab}         {\ensuremath{y_{\rm lab}}\xspace}
\newcommand{\etarange}[1]  {\mbox{$\left | \eta \right |~<~#1$}}
\newcommand{\yrange}[1]    {\mbox{$\left | y \right |~<~#1$}}
\newcommand{\dndy}         {\ensuremath{\mathrm{d}N_\mathrm{ch}/\mathrm{d}y}\xspace}
\newcommand{\dndeta}       {\ensuremath{\mathrm{d}N_\mathrm{ch}/\mathrm{d}\eta}\xspace}
\newcommand{\avdndeta}     {\ensuremath{\langle\dndeta\rangle}\xspace}
\newcommand{\dNdy}         {\ensuremath{\mathrm{d}N_\mathrm{ch}/\mathrm{d}y}\xspace}
\newcommand{\Npart}        {\ensuremath{N_\mathrm{part}}\xspace}
\newcommand{\Ncoll}        {\ensuremath{N_\mathrm{coll}}\xspace}
\newcommand{\dEdx}         {\ensuremath{\textrm{d}E/\textrm{d}x}\xspace}
\newcommand{\RpPb}         {\ensuremath{R_{\rm pPb}}\xspace}

\newcommand{\nineH}        {$\sqrt{s}~=~0.9$~Te\kern-.1emV\xspace}
\newcommand{\seven}        {$\sqrt{s}~=~7$~Te\kern-.1emV\xspace}
\newcommand{\twoH}         {$\sqrt{s}~=~0.2$~Te\kern-.1emV\xspace}
\newcommand{\twosevensix}  {$\sqrt{s}~=~2.76$~Te\kern-.1emV\xspace}
\newcommand{\five}         {$\sqrt{s}~=~5.02$~Te\kern-.1emV\xspace}
\newcommand{\twosevensixnn}{$\sqrt{s_{\mathrm{NN}}}~=~2.76$~Te\kern-.1emV\xspace}
\newcommand{\fivenn}       {$\sqrt{s_{\mathrm{NN}}}~=~5.02$~Te\kern-.1emV\xspace}
\newcommand{\LT}           {L{\'e}vy-Tsallis\xspace}
\newcommand{\GeVc}         {Ge\kern-.1emV/$c$\xspace}
\newcommand{\MeVc}         {Me\kern-.1emV/$c$\xspace}
\newcommand{\TeV}          {Te\kern-.1emV\xspace}
\newcommand{\GeV}          {Ge\kern-.1emV\xspace}
\newcommand{\MeV}          {Me\kern-.1emV\xspace}
\newcommand{\GeVmass}      {Ge\kern-.2emV/$c^2$\xspace}
\newcommand{\MeVmass}      {Me\kern-.2emV/$c^2$\xspace}
\newcommand{\lumi}         {\ensuremath{\mathcal{L}}\xspace}

\newcommand{\ITS}          {\rm{ITS}\xspace}
\newcommand{\TOF}          {\rm{TOF}\xspace}
\newcommand{\ZDC}          {\rm{ZDC}\xspace}
\newcommand{\ZDCs}         {\rm{ZDCs}\xspace}
\newcommand{\ZNA}          {\rm{ZNA}\xspace}
\newcommand{\ZNC}          {\rm{ZNC}\xspace}
\newcommand{\SPD}          {\rm{SPD}\xspace}
\newcommand{\SDD}          {\rm{SDD}\xspace}
\newcommand{\SSD}          {\rm{SSD}\xspace}
\newcommand{\TPC}          {\rm{TPC}\xspace}
\newcommand{\TRD}          {\rm{TRD}\xspace}
\newcommand{\VZERO}        {\rm{V0}\xspace}
\newcommand{\VZEROA}       {\rm{V0A}\xspace}
\newcommand{\VZEROC}       {\rm{V0C}\xspace}
\newcommand{\Vdecay} 	   {\ensuremath{V^{0}}\xspace}

\newcommand{\ee}           {\ensuremath{e^{+}e^{-}}} 
\newcommand{\pip}          {\ensuremath{\pi^{+}}\xspace}
\newcommand{\pim}          {\ensuremath{\pi^{-}}\xspace}
\newcommand{\kap}          {\ensuremath{\rm{K}^{+}}\xspace}
\newcommand{\kam}          {\ensuremath{\rm{K}^{-}}\xspace}
\newcommand{\pbar}         {\ensuremath{\rm\overline{p}}\xspace}
\newcommand{\kzero}        {\ensuremath{{\rm K}^{0}_{\rm{S}}}\xspace}
\newcommand{\lmb}          {\ensuremath{\Lambda}\xspace}
\newcommand{\almb}         {\ensuremath{\overline{\Lambda}}\xspace}
\newcommand{\Om}           {\ensuremath{\Omega^-}\xspace}
\newcommand{\Mo}           {\ensuremath{\overline{\Omega}^+}\xspace}
\newcommand{\X}            {\ensuremath{\Xi^-}\xspace}
\newcommand{\Ix}           {\ensuremath{\overline{\Xi}^+}\xspace}
\newcommand{\Xis}          {\ensuremath{\Xi^{\pm}}\xspace}
\newcommand{\Oms}          {\ensuremath{\Omega^{\pm}}\xspace}
\newcommand{\degree}       {\ensuremath{^{\rm o}}\xspace}

%% file: fa_2019-03-29.tex

The ALICE Collaboration would like to thank all its engineers and technicians for their invaluable contributions to the construction of the experiment and the CERN accelerator teams for the outstanding performance of the LHC complex.
The ALICE Collaboration gratefully acknowledges the resources and support provided by all Grid centres and the Worldwide LHC Computing Grid (WLCG) collaboration.
The ALICE Collaboration acknowledges the following funding agencies for their support in building and running the ALICE detector:
A. I. Alikhanyan National Science Laboratory (Yerevan Physics Institute) Foundation (ANSL), State Committee of Science and World Federation of Scientists (WFS), Armenia;
Austrian Academy of Sciences, Austrian Science Fund (FWF): [M 2467-N36] and Nationalstiftung f\"{u}r Forschung, Technologie und Entwicklung, Austria;
Ministry of Communications and High Technologies, National Nuclear Research Center, Azerbaijan;
Conselho Nacional de Desenvolvimento Cient\'{\i}fico e Tecnol\'{o}gico (CNPq), Universidade Federal do Rio Grande do Sul (UFRGS), Financiadora de Estudos e Projetos (Finep) and Funda\c{c}\~{a}o de Amparo \`{a} Pesquisa do Estado de S\~{a}o Paulo (FAPESP), Brazil;
Ministry of Science \& Technology of China (MSTC), National Natural Science Foundation of China (NSFC) and Ministry of Education of China (MOEC) , China;
Croatian Science Foundation and Ministry of Science and Education, Croatia;
Centro de Aplicaciones Tecnol\'{o}gicas y Desarrollo Nuclear (CEADEN), Cubaenerg\'{\i}a, Cuba;
Ministry of Education, Youth and Sports of the Czech Republic, Czech Republic;
The Danish Council for Independent Research | Natural Sciences, the Carlsberg Foundation and Danish National Research Foundation (DNRF), Denmark;
Helsinki Institute of Physics (HIP), Finland;
Commissariat \`{a} l'Energie Atomique (CEA), Institut National de Physique Nucl\'{e}aire et de Physique des Particules (IN2P3) and Centre National de la Recherche Scientifique (CNRS) and Rl\'{e}gion des  Pays de la Loire, France;
Bundesministerium f\"{u}r Bildung, Wissenschaft, Forschung und Technologie (BMBF) and GSI Helmholtzzentrum f\"{u}r Schwerionenforschung GmbH, Germany;
General Secretariat for Research and Technology, Ministry of Education, Research and Religions, Greece;
National Research, Development and Innovation Office, Hungary;
Department of Atomic Energy Government of India (DAE), Department of Science and Technology, Government of India (DST), University Grants Commission, Government of India (UGC) and Council of Scientific and Industrial Research (CSIR), India;
Indonesian Institute of Science, Indonesia;
Centro Fermi - Museo Storico della Fisica e Centro Studi e Ricerche Enrico Fermi and Istituto Nazionale di Fisica Nucleare (INFN), Italy;
Institute for Innovative Science and Technology , Nagasaki Institute of Applied Science (IIST), Japan Society for the Promotion of Science (JSPS) KAKENHI and Japanese Ministry of Education, Culture, Sports, Science and Technology (MEXT), Japan;
Consejo Nacional de Ciencia (CONACYT) y Tecnolog\'{i}a, through Fondo de Cooperaci\'{o}n Internacional en Ciencia y Tecnolog\'{i}a (FONCICYT) and Direcci\'{o}n General de Asuntos del Personal Academico (DGAPA), Mexico;
Nederlandse Organisatie voor Wetenschappelijk Onderzoek (NWO), Netherlands;
The Research Council of Norway, Norway;
Commission on Science and Technology for Sustainable Development in the South (COMSATS), Pakistan;
Pontificia Universidad Cat\'{o}lica del Per\'{u}, Peru;
Ministry of Science and Higher Education and National Science Centre, Poland;
Korea Institute of Science and Technology Information and National Research Foundation of Korea (NRF), Republic of Korea;
Ministry of Education and Scientific Research, Institute of Atomic Physics and Ministry of Research and Innovation and Institute of Atomic Physics, Romania;
Joint Institute for Nuclear Research (JINR), Ministry of Education and Science of the Russian Federation, National Research Centre Kurchatov Institute, Russian Science Foundation and Russian Foundation for Basic Research, Russia;
Ministry of Education, Science, Research and Sport of the Slovak Republic, Slovakia;
National Research Foundation of South Africa, South Africa;
Swedish Research Council (VR) and Knut \& Alice Wallenberg Foundation (KAW), Sweden;
European Organization for Nuclear Research, Switzerland;
National Science and Technology Development Agency (NSDTA), Suranaree University of Technology (SUT) and Office of the Higher Education Commission under NRU project of Thailand, Thailand;
Turkish Atomic Energy Agency (TAEK), Turkey;
National Academy of  Sciences of Ukraine, Ukraine;
Science and Technology Facilities Council (STFC), United Kingdom;
National Science Foundation of the United States of America (NSF) and United States Department of Energy, Office of Nuclear Physics (DOE NP), United States of America.

%% file: 2019-03-29-Alice_Authorlist_2019-Mar-29.tex

\begingroup
\small
\begin{flushleft}
S.~Acharya\Irefn{org141}\And 
D.~Adamov\'{a}\Irefn{org93}\And 
S.P.~Adhya\Irefn{org141}\And 
A.~Adler\Irefn{org74}\And 
J.~Adolfsson\Irefn{org80}\And 
M.M.~Aggarwal\Irefn{org98}\And 
G.~Aglieri Rinella\Irefn{org34}\And 
M.~Agnello\Irefn{org31}\And 
N.~Agrawal\Irefn{org10}\And 
Z.~Ahammed\Irefn{org141}
% ...
\And
\renewcommand\labelenumi{\textsuperscript{\theenumi}~}

\section*{Affiliation notes}
\renewcommand\theenumi{\roman{enumi}}
\begin{Authlist}
\item \Adef{org*}Deceased
\item \Adef{orgI}Dipartimento DET del Politecnico di Torino, Turin, Italy
\item \Adef{orgII}M.V. Lomonosov Moscow State University, D.V. Skobeltsyn Institute of Nuclear, Physics, Moscow, Russia
\item \Adef{orgIII}Department of Applied Physics, Aligarh Muslim University, Aligarh, India
\item \Adef{orgIV}Institute of Theoretical Physics, University of Wroclaw, Poland
\end{Authlist}

\section*{Collaboration Institutes}
\renewcommand\theenumi{\arabic{enumi}~}
\begin{Authlist}
\item \Idef{org1}A.I. Alikhanyan National Science Laboratory (Yerevan Physics Institute) Foundation, Yerevan, Armenia
\item \Idef{org2}Bogolyubov Institute for Theoretical Physics, National Academy of Sciences of Ukraine, Kiev, Ukraine
\item \Idef{org3}Bose Institute, Department of Physics  and Centre for Astroparticle Physics and Space Science (CAPSS), Kolkata, India
\item \Idef{org4}Budker Institute for Nuclear Physics, Novosibirsk, Russia
\item \Idef{org5}California Polytechnic State University, San Luis Obispo, California, United States
\item \Idef{org6}Central China Normal University, Wuhan, China
\item \Idef{org7}Centre de Calcul de l'IN2P3, Villeurbanne, Lyon, France
\item \Idef{org8}Centro de Aplicaciones Tecnol\'{o}gicas y Desarrollo Nuclear (CEADEN), Havana, Cuba
\item \Idef{org9}Centro de Investigaci\'{o}n y de Estudios Avanzados (CINVESTAV), Mexico City and M\'{e}rida, Mexico
\item \Idef{org10}Centro Fermi - Museo Storico della Fisica e Centro Studi e Ricerche ``Enrico Fermi', Rome, Italy
\item \Idef{org11}Chicago State University, Chicago, Illinois, United States
\item \Idef{org12}China Institute of Atomic Energy, Beijing, China
\item \Idef{org13}Chonbuk National University, Jeonju, Republic of Korea
\item \Idef{org14}Comenius University Bratislava, Faculty of Mathematics, Physics and Informatics, Bratislava, Slovakia
\item \Idef{org15}COMSATS University Islamabad, Islamabad, Pakistan
\item \Idef{org16}Creighton University, Omaha, Nebraska, United States
\item \Idef{org17}Department of Physics, Aligarh Muslim University, Aligarh, India
\item \Idef{org18}Department of Physics, Pusan National University, Pusan, Republic of Korea
\item \Idef{org19}Department of Physics, Sejong University, Seoul, Republic of Korea
\item \Idef{org20}Department of Physics, University of California, Berkeley, California, United States
\item \Idef{org21}Department of Physics, University of Oslo, Oslo, Norway
\item \Idef{org22}Department of Physics and Technology, University of Bergen, Bergen, Norway
\item \Idef{org23}Dipartimento di Fisica dell'Universit\`{a} 'La Sapienza' and Sezione INFN, Rome, Italy
\item \Idef{org24}Dipartimento di Fisica dell'Universit\`{a} and Sezione INFN, Cagliari, Italy
\item \Idef{org25}Dipartimento di Fisica dell'Universit\`{a} and Sezione INFN, Trieste, Italy
\item \Idef{org26}Dipartimento di Fisica dell'Universit\`{a} and Sezione INFN, Turin, Italy
\item \Idef{org27}Dipartimento di Fisica e Astronomia dell'Universit\`{a} and Sezione INFN, Bologna, Italy
\item \Idef{org28}Dipartimento di Fisica e Astronomia dell'Universit\`{a} and Sezione INFN, Catania, Italy
\item \Idef{org29}Dipartimento di Fisica e Astronomia dell'Universit\`{a} and Sezione INFN, Padova, Italy
\item \Idef{org30}Dipartimento di Fisica `E.R.~Caianiello' dell'Universit\`{a} and Gruppo Collegato INFN, Salerno, Italy
\item \Idef{org31}Dipartimento DISAT del Politecnico and Sezione INFN, Turin, Italy
\item \Idef{org32}Dipartimento di Scienze e Innovazione Tecnologica dell'Universit\`{a} del Piemonte Orientale and INFN Sezione di Torino, Alessandria, Italy
\item \Idef{org33}Dipartimento Interateneo di Fisica `M.~Merlin' and Sezione INFN, Bari, Italy
\item \Idef{org34}European Organization for Nuclear Research (CERN), Geneva, Switzerland
\item \Idef{org35}Faculty of Electrical Engineering, Mechanical Engineering and Naval Architecture, University of Split, Split, Croatia
\item \Idef{org36}Faculty of Engineering and Science, Western Norway University of Applied Sciences, Bergen, Norway
\item \Idef{org37}Faculty of Nuclear Sciences and Physical Engineering, Czech Technical University in Prague, Prague, Czech Republic
\item \Idef{org38}Faculty of Science, P.J.~\v{S}af\'{a}rik University, Ko\v{s}ice, Slovakia
\item \Idef{org39}Frankfurt Institute for Advanced Studies, Johann Wolfgang Goethe-Universit\"{a}t Frankfurt, Frankfurt, Germany
\item \Idef{org40}Gangneung-Wonju National University, Gangneung, Republic of Korea
\item \Idef{org41}Gauhati University, Department of Physics, Guwahati, India
\item \Idef{org42}Helmholtz-Institut f\"{u}r Strahlen- und Kernphysik, Rheinische Friedrich-Wilhelms-Universit\"{a}t Bonn, Bonn, Germany
\item \Idef{org43}Helsinki Institute of Physics (HIP), Helsinki, Finland
\item \Idef{org44}High Energy Physics Group,  Universidad Aut\'{o}noma de Puebla, Puebla, Mexico
\item \Idef{org45}Hiroshima University, Hiroshima, Japan
\item \Idef{org46}Hochschule Worms, Zentrum  f\"{u}r Technologietransfer und Telekommunikation (ZTT), Worms, Germany
\item \Idef{org47}Horia Hulubei National Institute of Physics and Nuclear Engineering, Bucharest, Romania
\item \Idef{org48}Indian Institute of Technology Bombay (IIT), Mumbai, India
\item \Idef{org49}Indian Institute of Technology Indore, Indore, India
\item \Idef{org50}Indonesian Institute of Sciences, Jakarta, Indonesia
\item \Idef{org51}INFN, Laboratori Nazionali di Frascati, Frascati, Italy
\item \Idef{org52}INFN, Sezione di Bari, Bari, Italy
\item \Idef{org53}INFN, Sezione di Bologna, Bologna, Italy
\item \Idef{org54}INFN, Sezione di Cagliari, Cagliari, Italy
\item \Idef{org55}INFN, Sezione di Catania, Catania, Italy
\item \Idef{org56}INFN, Sezione di Padova, Padova, Italy
\item \Idef{org57}INFN, Sezione di Roma, Rome, Italy
\item \Idef{org58}INFN, Sezione di Torino, Turin, Italy
\item \Idef{org59}INFN, Sezione di Trieste, Trieste, Italy
\item \Idef{org60}Inha University, Incheon, Republic of Korea
\item \Idef{org61}Institut de Physique Nucl\'{e}aire d'Orsay (IPNO), Institut National de Physique Nucl\'{e}aire et de Physique des Particules (IN2P3/CNRS), Universit\'{e} de Paris-Sud, Universit\'{e} Paris-Saclay, Orsay, France
\item \Idef{org62}Institute for Nuclear Research, Academy of Sciences, Moscow, Russia
\item \Idef{org63}Institute for Subatomic Physics, Utrecht University/Nikhef, Utrecht, Netherlands
\item \Idef{org64}Institute for Theoretical and Experimental Physics, Moscow, Russia
\item \Idef{org65}Institute of Experimental Physics, Slovak Academy of Sciences, Ko\v{s}ice, Slovakia
\item \Idef{org66}Institute of Physics, Homi Bhabha National Institute, Bhubaneswar, India
\item \Idef{org67}Institute of Physics of the Czech Academy of Sciences, Prague, Czech Republic
\item \Idef{org68}Institute of Space Science (ISS), Bucharest, Romania
\item \Idef{org69}Institut f\"{u}r Kernphysik, Johann Wolfgang Goethe-Universit\"{a}t Frankfurt, Frankfurt, Germany
\item \Idef{org70}Instituto de Ciencias Nucleares, Universidad Nacional Aut\'{o}noma de M\'{e}xico, Mexico City, Mexico
\item \Idef{org71}Instituto de F\'{i}sica, Universidade Federal do Rio Grande do Sul (UFRGS), Porto Alegre, Brazil
\item \Idef{org72}Instituto de F\'{\i}sica, Universidad Nacional Aut\'{o}noma de M\'{e}xico, Mexico City, Mexico
\item \Idef{org73}iThemba LABS, National Research Foundation, Somerset West, South Africa
\item \Idef{org74}Johann-Wolfgang-Goethe Universit\"{a}t Frankfurt Institut f\"{u}r Informatik, Fachbereich Informatik und Mathematik, Frankfurt, Germany
\item \Idef{org75}Joint Institute for Nuclear Research (JINR), Dubna, Russia
\item \Idef{org76}Korea Institute of Science and Technology Information, Daejeon, Republic of Korea
\item \Idef{org77}KTO Karatay University, Konya, Turkey
\item \Idef{org78}Laboratoire de Physique Subatomique et de Cosmologie, Universit\'{e} Grenoble-Alpes, CNRS-IN2P3, Grenoble, France
\item \Idef{org79}Lawrence Berkeley National Laboratory, Berkeley, California, United States
\item \Idef{org80}Lund University Department of Physics, Division of Particle Physics, Lund, Sweden
\item \Idef{org81}Nagasaki Institute of Applied Science, Nagasaki, Japan
\item \Idef{org82}Nara Women{'}s University (NWU), Nara, Japan
\item \Idef{org83}National and Kapodistrian University of Athens, School of Science, Department of Physics , Athens, Greece
\item \Idef{org84}National Centre for Nuclear Research, Warsaw, Poland
\item \Idef{org85}National Institute of Science Education and Research, Homi Bhabha National Institute, Jatni, India
\item \Idef{org86}National Nuclear Research Center, Baku, Azerbaijan
\item \Idef{org87}National Research Centre Kurchatov Institute, Moscow, Russia
\item \Idef{org88}Niels Bohr Institute, University of Copenhagen, Copenhagen, Denmark
\item \Idef{org89}Nikhef, National institute for subatomic physics, Amsterdam, Netherlands
\item \Idef{org90}NRC Kurchatov Institute IHEP, Protvino, Russia
\item \Idef{org91}NRNU Moscow Engineering Physics Institute, Moscow, Russia
\item \Idef{org92}Nuclear Physics Group, STFC Daresbury Laboratory, Daresbury, United Kingdom
\item \Idef{org93}Nuclear Physics Institute of the Czech Academy of Sciences, \v{R}e\v{z} u Prahy, Czech Republic
\item \Idef{org94}Oak Ridge National Laboratory, Oak Ridge, Tennessee, United States
\item \Idef{org95}Ohio State University, Columbus, Ohio, United States
\item \Idef{org96}Petersburg Nuclear Physics Institute, Gatchina, Russia
\item \Idef{org97}Physics department, Faculty of science, University of Zagreb, Zagreb, Croatia
\item \Idef{org98}Physics Department, Panjab University, Chandigarh, India
\item \Idef{org99}Physics Department, University of Jammu, Jammu, India
\item \Idef{org100}Physics Department, University of Rajasthan, Jaipur, India
\item \Idef{org101}Physikalisches Institut, Eberhard-Karls-Universit\"{a}t T\"{u}bingen, T\"{u}bingen, Germany
\item \Idef{org102}Physikalisches Institut, Ruprecht-Karls-Universit\"{a}t Heidelberg, Heidelberg, Germany
\item \Idef{org103}Physik Department, Technische Universit\"{a}t M\"{u}nchen, Munich, Germany
\item \Idef{org104}Politecnico di Bari, Bari, Italy
\item \Idef{org105}Research Division and ExtreMe Matter Institute EMMI, GSI Helmholtzzentrum f\"ur Schwerionenforschung GmbH, Darmstadt, Germany
\item \Idef{org106}Rudjer Bo\v{s}kovi\'{c} Institute, Zagreb, Croatia
\item \Idef{org107}Russian Federal Nuclear Center (VNIIEF), Sarov, Russia
\item \Idef{org108}Saha Institute of Nuclear Physics, Homi Bhabha National Institute, Kolkata, India
\item \Idef{org109}School of Physics and Astronomy, University of Birmingham, Birmingham, United Kingdom
\item \Idef{org110}Secci\'{o}n F\'{\i}sica, Departamento de Ciencias, Pontificia Universidad Cat\'{o}lica del Per\'{u}, Lima, Peru
\item \Idef{org111}Shanghai Institute of Applied Physics, Shanghai, China
\item \Idef{org112}St. Petersburg State University, St. Petersburg, Russia
\item \Idef{org113}Stefan Meyer Institut f\"{u}r Subatomare Physik (SMI), Vienna, Austria
\item \Idef{org114}SUBATECH, IMT Atlantique, Universit\'{e} de Nantes, CNRS-IN2P3, Nantes, France
\item \Idef{org115}Suranaree University of Technology, Nakhon Ratchasima, Thailand
\item \Idef{org116}Technical University of Ko\v{s}ice, Ko\v{s}ice, Slovakia
\item \Idef{org117}Technische Universit\"{a}t M\"{u}nchen, Excellence Cluster 'Universe', Munich, Germany
\item \Idef{org118}The Henryk Niewodniczanski Institute of Nuclear Physics, Polish Academy of Sciences, Cracow, Poland
\item \Idef{org119}The University of Texas at Austin, Austin, Texas, United States
\item \Idef{org120}Universidad Aut\'{o}noma de Sinaloa, Culiac\'{a}n, Mexico
\item \Idef{org121}Universidade de S\~{a}o Paulo (USP), S\~{a}o Paulo, Brazil
\item \Idef{org122}Universidade Estadual de Campinas (UNICAMP), Campinas, Brazil
\item \Idef{org123}Universidade Federal do ABC, Santo Andre, Brazil
\item \Idef{org124}University College of Southeast Norway, Tonsberg, Norway
\item \Idef{org125}University of Cape Town, Cape Town, South Africa
\item \Idef{org126}University of Houston, Houston, Texas, United States
\item \Idef{org127}University of Jyv\"{a}skyl\"{a}, Jyv\"{a}skyl\"{a}, Finland
\item \Idef{org128}University of Liverpool, Liverpool, United Kingdom
\item \Idef{org129}University of Science and Techonology of China, Hefei, China
\item \Idef{org130}University of Tennessee, Knoxville, Tennessee, United States
\item \Idef{org131}University of the Witwatersrand, Johannesburg, South Africa
\item \Idef{org132}University of Tokyo, Tokyo, Japan
\item \Idef{org133}University of Tsukuba, Tsukuba, Japan
\item \Idef{org134}Universit\'{e} Clermont Auvergne, CNRS/IN2P3, LPC, Clermont-Ferrand, France
\item \Idef{org135}Universit\'{e} de Lyon, Universit\'{e} Lyon 1, CNRS/IN2P3, IPN-Lyon, Villeurbanne, Lyon, France
\item \Idef{org136}Universit\'{e} de Strasbourg, CNRS, IPHC UMR 7178, F-67000 Strasbourg, France, Strasbourg, France
\item \Idef{org137}Universit\'{e} Paris-Saclay Centre d'Etudes de Saclay (CEA), IRFU, D\'{e}partment de Physique Nucl\'{e}aire (DPhN), Saclay, France
\item \Idef{org138}Universit\`{a} degli Studi di Foggia, Foggia, Italy
\item \Idef{org139}Universit\`{a} degli Studi di Pavia, Pavia, Italy
\item \Idef{org140}Universit\`{a} di Brescia, Brescia, Italy
\item \Idef{org141}Variable Energy Cyclotron Centre, Homi Bhabha National Institute, Kolkata, India
\item \Idef{org142}Warsaw University of Technology, Warsaw, Poland
\item \Idef{org143}Wayne State University, Detroit, Michigan, United States
\item \Idef{org144}Westf\"{a}lische Wilhelms-Universit\"{a}t M\"{u}nster, Institut f\"{u}r Kernphysik, M\"{u}nster, Germany
\item \Idef{org145}Wigner Research Centre for Physics, Hungarian Academy of Sciences, Budapest, Hungary
\item \Idef{org146}Yale University, New Haven, Connecticut, United States
\item \Idef{org147}Yonsei University, Seoul, Republic of Korea
\end{Authlist}
\endgroup